\documentclass{aa}
\usepackage[varg]{txfonts}
\usepackage{graphicx}
\usepackage[colorlinks, urlcolor = blue, linkcolor = blue, citecolor = blue]{hyperref}
\bibpunct{(}{)}{;}{a}{}{,} % to follow the A&A style

\usepackage{placeins}
\usepackage{multicol}
\usepackage{capt-of}
\usepackage[caption=false]{subfig}
\usepackage{stfloats}
\usepackage{gensymb}
\usepackage[version=4]{mhchem}
\usepackage{textcomp}
\usepackage{ctable}
\usepackage{csquotes}
\usepackage[toc,page]{appendix}
\title{
    Dynamics and potential origins of decimeter-sized particles around comet 67P/Churyumov-Gerasimenko
}
\titlerunning{Dynamics and potential origins of decimeter-sized particles around comet 67P/Churyumov-Gerasimenko}

\author{
    Marius Pfeifer\inst{1}
    \and Jessica Agarwal\inst{1, 2}
    \and Raphael Marschall\inst{3}
    \and Björn Grieger\inst{4}
    \and Pablo Lemos\inst{1, 2}
}

\institute{
    Max Planck Institute for Solar System Research, Justus-von-Liebig-Weg 3, D-37077 Göttingen, Germany\\
    \email{\href{mailto:pfeifer@mps.mpg.de}{pfeifer@mps.mpg.de}}
    \and Institute for Geophysics and Extraterrestrial Physics, TU Braunschweig, Mendelssohnstraße 3, D-38106 Braunschweig, Germany
    \and CNRS, Laboratoire J.-L. Lagrange, Observatoire de la Côte d’Azur, Nice, France
    \and Aurora Technology B.V. for ESA, ESAC, Madrid, Spain
}

\date{Received <date> / Accepted <date>}

\abstract{One of the primary goals of the European Space Agency's Rosetta mission to comet 67P/Churyumov-Gerasimenko was to investigate the mechanisms responsible for cometary activity. 
}{Our aim is to learn more about the ejection process of large refractory material by studying the dynamics of decimeter-sized dust particles in the coma of 67P and estimating their potential source regions.
}{We algorithmically tracked thousands of individual particles through four OSIRIS/NAC image sequences of 67P's near-nucleus coma. We then traced concentrated particle groups back to the nucleus surface, and estimated their potential source regions, size distributions, and projected dynamical parameters. Finally, we compared the observed activity to dust coma simulations.
}{We traced back 409 decimeter-sized particles to four suspected source regions. The regions strongly overlap and are mostly confined to the Khonsu-Atum-Anubis area. The activity may be linked to rugged terrain, and the erosion of fine dust and the ejection of large boulders may be mutually exclusive. Power-law indices fitted to the particle size--frequency distributions range from $3.4 \pm 0.3$ to $3.8 \pm 0.4$. Gas drag fits to the radial particle accelerations provide an estimate for the local gas production rates (\mbox{$Q_\text{g} = 3.6 \cdot 10^{-5}$\,kg\,s$^{-1}$\,m$^{-2}$}), which is several times higher than our model predictions based on purely insolation-driven water ice sublimation. Our observational results and our modeling results both reveal that our particles were likely ejected with substantial nonzero initial velocities of around 0.5--0.6\,m\,s$^{-1}$.
}{Our findings strongly suggest that the observed ejection of decimeter-sized particles cannot be explained by water ice sublimation and favorable illumination conditions alone. Instead, the local structures and compositions of the source regions likely play a major role. In line with current ejection models of decimeter-sized particles, we deem an overabundance of \ce{CO2} ice and its sublimation to be the most probable driver. In addition, because of the significant initial velocities, we suspect the ejection events to be considerably more energetic than gradual liftoffs. 
}


\begin{document}
    \maketitle

    \section{Introduction}
    \label{sect:introduction}

        Comets are small Solar System objects that formed in the outer regions of our protoplanetary disk beyond the snowline \citep[e.g.,][]{weissmanOriginEvolutionCometary2020}. At these heliocentric distances, water and other (super-)volatiles like \ce{CO2} can remain solid over astronomical timescales. These ices make up a significant part of cometary material, which otherwise consists mainly of refractory aggregates. Because of this ice content, comets become active once they enter the inner Solar System: their ices start to sublimate. If this happens beneath the cometary surface, the expanding gas can get trapped and build up pressure. Eventually, this pressure may overcome the gravity and tensile strength of the overlying material, expel the gas, and eject some of the refractory material along with it. The released gas and dust then form the characteristic cometary coma, tail, and trail \citep[e.g.,][]{zakharovPhysicalProcessesLeading2022}.
        
        Decimeter-sized particles\footnote{Following the classification of cometary dust by \citet{guttlerSynthesisMorphologicalDescription2019}, we use the term \enquote{particle} {as a generic term for any unspecified dust particle}, independently of its size.} are likely ejected by \ce{CO2} ice sublimation in deeper surface layers \citep[e.g.,][]{gundlachActivityCometsUnderstanding2020, fulleHowCometsWork2020, wesolowskiSelectedMechanismsMatter2020, ciarnielloMacroMicroStructures2022, davidssonCO2drivenSurfaceChanges2022}; however, the responsible mechanisms are not yet fully understood (e.g., \citealt{zakharovPhysicalProcessesLeading2022, bischoffQuantitativeDescriptionComet2023, agarwalDustEmissionDynamics2023}), and even less so for small particles where cohesion forces dominate (e.g., \citealt{gundlachWhatDrivesDust2015, skorovNearsurfaceIceDriver2017, markkanenThermophysicalModelIcy2020}). 
        
        Learning more about these processes was, and still is, one of the primary science goals of the European Space Agency's Rosetta mission to comet \object{67P/Churyumov–Gerasimenko} \citep[e.g.,][]{taylorRosettaMissionOrbiter2017}. The spacecraft rendezvoused with 67P in August 2014 and accompanied the comet through its perihelion passage in August 2015, until Rosetta was set on an intercept course with 67P on September 30, 2016, and landed on its surface. Among the suite of Rosetta's science instruments was the Optical, Spectroscopic, and Infrared Remote Imaging System (OSIRIS), which consisted of a wide-angle and a narrow-angle camera \citep[WAC and NAC,][]{kellerOSIRISScientificCamera2007}. During the rendezvous phase, these cameras recorded many image sequences of the near-nucleus coma, which also captured the motion of individual dust particles shortly after their ejection. Because the study of their dynamics can help to understand their ejection process, we developed a particle tracking algorithm for OSIRIS images in \citet{pfeiferTrailCometTail2022}. 
        
        For the current paper we have now applied an enhanced version of this algorithm to four OSIRIS/NAC image sequences. In Section~\ref{sect:data_selection} we give an overview of the image sequences. In Section~\ref{sect:particle_tracking} we briefly discuss how we improved the tracking algorithm, and explain why we only focus on certain particle groups for our analysis, and how we selected them and identified their suspected source regions. In Section~\ref{sect:results_and_discussion} we investigate these regions, determine the particle size--frequency distributions, reconstruct the local illumination conditions and surface accelerations, analyze the particle dynamics, and compare the results to dust coma simulations. Finally, in Section~\ref{sect:summary_and_conclusions} we summarize our findings and conclusions.

    \section{Data selection}
    \label{sect:data_selection}

        We chose the four most suitable image sequences for our analysis from a group of about 100 similar sequences. The others were rejected for various reasons (listed in Appendix~\ref{sect:rejection_reasons}), but to summarize, we looked for (preferably complete) image sequences that were recorded in image pairs over a roughly two-hour time span and no farther than $\approx 150$\,km from the nucleus, and that consist of two subsequences (one short, one long), contain (trackable) sidereal objects, and show at least one concentrated group of high-quality particle tracks.
            
\begin{figure}[!ht]
            \centering
            \includegraphics[width=\linewidth]{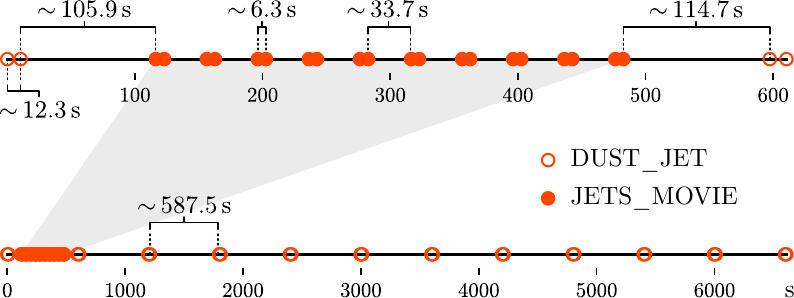}
            \caption{Typical timeline of the four image sequences analyzed, in this case STP089. The other sequences deviate from this structure only by a few seconds at most.}
            \label{fig:STP089_timeline}
        \end{figure}

\begin{figure*}[!ht]
            \centering
            \includegraphics[width=\linewidth]{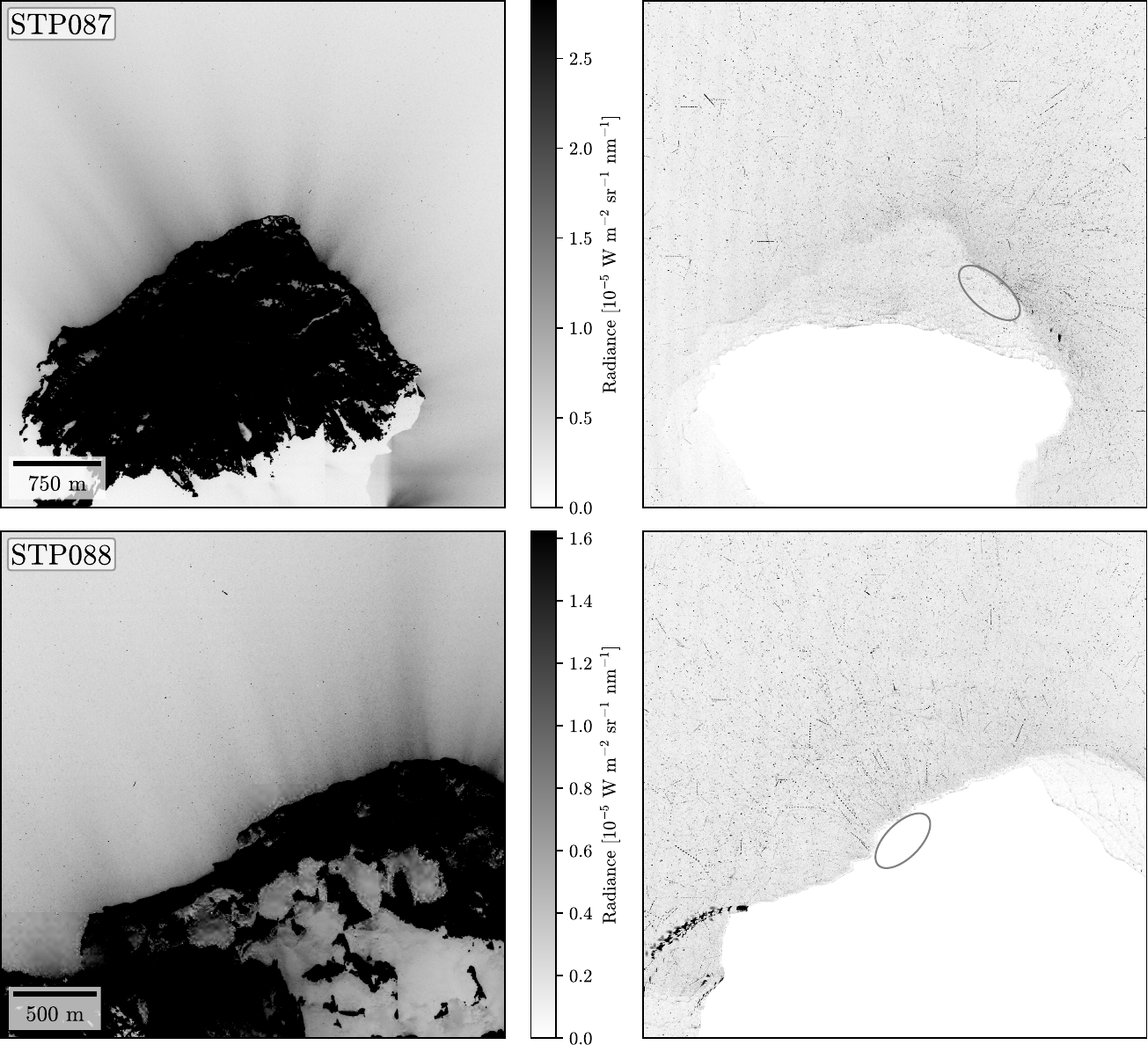}
            \caption{Sample images from sequences STP087 and STP088. First images of the respective sequences on the left, master images on the right. The ellipses in the master images mark the suspected source regions of the concentrated particle groups. All images are brightness-inverted and had their contrasts improved for better readability (for sequences STP089 and STP090 see Fig.~\ref{fig:sample_images_stp089_stp090} in the Appendix).}
            \label{fig:sample_images_stp087_stp088}
        \end{figure*}

        The four selected sequences (see sample images in Figs.~\ref{fig:sample_images_stp087_stp088} and \ref{fig:sample_images_stp089_stp090}) were recorded by OSIRIS/NAC between December 16, 2015, and January 6, 2016 \citep[see Table~\ref{tab:nac_specs} and][]{kellerOSIRISScientificCamera2007}. They each consist of 44 images and are the result of merging two subsequences (OSIRIS activity tags “JETS\_MOVIE” and “DUST\_JET”). Figure~\ref{fig:STP089_timeline} shows the typical timeline of such a sequence, and Table~\ref{tab:meta_data} lists relevant metadata \citep[see also][]{pfeiferTrailCometTail2022}. In the following, we call these sequences according to their designated (Rosetta mission) short-time-planning (STP) numbers: STP087, STP088, STP089, and STP090. 
        
        \ctable[
            caption = OSIRIS/NAC specifications.,
            label   = tab:nac_specs,
            pos             = b,
            width   = \linewidth,
            notespar
        ]{lc}{
        }{                      \FL
            Field of view (FOV)     & $2.208\degree \times 2.208\degree$             \NN 
            Pixel resolution           & $18.6\,\text{\textmu rad} \times 18.6$\,\textmu rad \NN
            CCD resolution           & $2048\,\text{px} \times 2048$\,px                                \NN
            Filter (NAC F22)           & center: 649.2\,nm, bandwidth: 85\,nm     \LL
        }
        
        \ctable[
            caption = Metadata of the four sequences   analyzed.,
            label   = tab:meta_data,
            pos         = b,
            width   = \textwidth,
            notespar,
            star
        ]{lrrrrrrrrrrr}{
            \tnote[a]{Starting time.}
            \tnote[b]{Total duration.}
            \tnote[c]{Mean heliocentric distance.}
            \tnote[d]{Mean nucleocentric distance.}
            \tnote[e]{Mean pixel resolution at nucleocentric distance.}
            \tnote[f]{Mean CCD FOV at nucleocentric distance.}
            \tnote[g]{Mean sub-solar latitude.} 
            \tnote[h]{Sub-solar longitude range.}
            \tnote[i]{Mean phase angle.}
        }{                \FL  
            Name    & Date  & $t_0$\tmark[a] (UT)  & $\Delta t$\tmark[b] &   $\bar{r}_\text{h}$\tmark[c] [AU]        & $\bar{\Delta}$\tmark[d] [km] & $\bar{\delta}_\text{px}$\tmark[e] [m$^2$] & $\bar{\delta}_\text{FOV}$\tmark[f] [km$^2$] & $\bar{\phi}_\text{s}$\tmark[g] [$\degree$]  & $\lambda_\text{s}$\tmark[h] [$\degree$]  & $\bar{\alpha}$\tmark[i] [$\degree$]  \ML
            STP087 & 2015-12-16   & 7:01:06   &  1:50:10   & $1.90$   & $113$  & $2.1 \times 2.1$ & $4.35 \times 4.35$  & $-21.1$    & $122$--$177$   & $90.9$   \ML
            STP088 & 2015-12-26   & 6:30:06   &  1:50:11   & $1.98$   & $79$   & $1.5 \times 1.5$ & $3.06 \times 3.06$  & $-18.2$    & $185$--$240$   & $91.2$   \ML
            STP089 & 2015-12-30   & 7:01:06   &  1:50:10   & $2.01$   & $88$   & $1.7 \times 1.7$ & $3.39 \times 3.39$  & $-17.1$    & $188$--$243$   & $91.3$   \ML
            STP090 & 2016-01-06   & 7:01:04   &  1:50:11   & $2.06$   & $86$   & $1.6 \times 1.6$ & $3.33 \times 3.33$  & $-15.3$    & $219$--$274$   & $91.2$   \LL
        }
        
        As did \citet{pfeiferTrailCometTail2022}, we used images of OSIRIS calibration level 3E \citep{tubianaScientificAssessmentQuality2015}.\footnote{The data are available at the Planetary Science Archive of the European Space Agency under \url{https://www.cosmos.esa.int/web/psa/rosetta}.} Images of this level are radiometrically calibrated, corrected for in-field and solar stay-light, geometric distortion, and boresight offset (resampled), and are expressed in radiance units (W\,m$^{-2}$\,sr$^{-1}$\,nm$^{-1}$). In preparation for the particle tracking, the images additionally underwent a number of pre-processing steps, where the diffuse coma background was removed, the bright nucleus masked out, and the point-source-like particles were detected \citep{pfeiferTrailCometTail2022}. For visual confirmation, we then also stacked the cleaned images using only the maximum value that each pixel assumed over the course of the sequence to create a \enquote{master image} (see, e.g., Fig.~\ref{fig:sample_images_stp087_stp088}). The identified point source coordinates, on the other hand, are passed on to the tracking algorithm, where they are used to reconstruct the particle trajectories.
                
    \section{Particle tracking}
    \label{sect:particle_tracking}

        \subsection{Tracking algorithm}
        \label{sect:tracking_algorithm}
        
            \begin{figure*}[!ht]
                \centering
                \includegraphics[width=\linewidth]{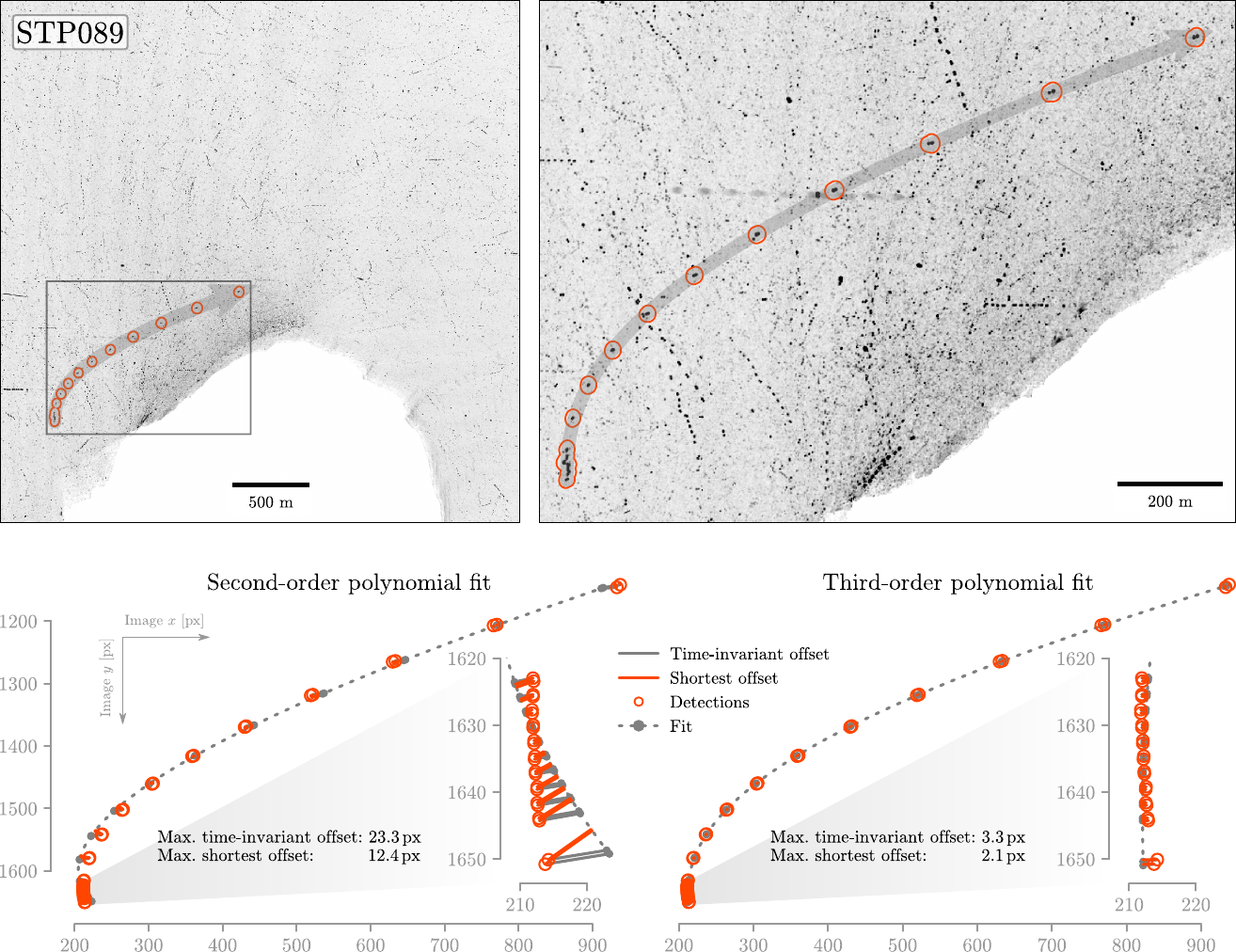}
                \caption{Sample track from sequence STP089 demonstrating how fitting third-order polynomials and calculating the residual as the shortest offset instead of the time-invariant offset can significantly improve the tracking results. Full (brightness-inverted) master image on the top left, close-up on the top  right. The sample track is highlighted by the gray arrow, while the orange contours indicate the corresponding point sources.}
                \label{fig:sample_track}
            \end{figure*} 
    
            To track the detected particles, we used the algorithm described in \citet[see this reference for definitions of technical terms used in the following]{pfeiferTrailCometTail2022}, which we improved in three significant ways: 
            
            1) Because we found that for some tracks, fitting a second-order polynomial is inadequate, we now also allow fitting of third-order polynomials during the extended tracking phase, but only under the condition that the absolute jerk (i.e., change in acceleration) is lower than a certain user-defined threshold. This condition is necessary because the jerk is very sensitive to inaccurate data points. Thus, high jerk values more likely result from pointing fluctuation that we could not completely correct for or from unrelated detections that were erroneously incorporated into a track, rather than from actual physical forces that acted on the particles. Because the tracks generally have a lot more data points than the polynomials have degrees of freedom, and because their fits yield median adjusted $R^2$ values \citep[a measure for the goodness of fit; e.g.,][]{mordecalezekielMethodsCorrelationAnalysis1930, fahrmeirRegressionModelsMethods2021} that are extremely close to unity ($\gtrapprox 0.9999$), we are confident that we are not overfitting our data. Nevertheless, there are other caveats that come with the choice of polynomial degree, specifically concerning the extrapolation of such fits (see Appendix~\ref{sect:fitting_caveats}). We therefore limit extrapolations to a maximum of 30 minutes.  
            
            2) Originally, some detections were rejected during the tracking process, even though they were located close to their track and could be visually confirmed to be part of it. We found that they were rejected because their residual offset was too large with respect to their expected location at the corresponding time, according to the fit. We now calculate the residual offset as the shortest distance to the fit, effectively dropping the timing criterion. 
            
            3) The radiance values of particle detections are now more accurate. Instead of integrating only over the member-pixels of a point source that are above a certain radiance threshold, we now integrate over all the (sub)pixels that lie within an ellipse containing most of the source's signal \citep[see also][]{bertinSExtractorSoftwareSource1996, barbarySEPSourceExtractor2016}. During the photometry however, we are deliberately not removing any background signal (e.g., derived from elliptical annuli around the detections), because we already removed the local background in preparation for the detection process, on a scale ($16 \times 16$\,px) very similar to typical annuli radii \citep[$\approx 11$--$17$px; see also][]{pfeiferTrailCometTail2022}. 

            Figure~\ref{fig:sample_track} shows how fitting third-order polynomials in certain cases, and generally calculating the residual offset as the shortest distance to the fit, improves the particle tracking. In the demonstrated case, fitting a third-order polynomial is even necessary to track the particle at all. If a second-order polynomial were used, the tracking parameters would have to be too lenient, which would result in highly inaccurate velocity and acceleration vectors and thus predictions during tracking, causing the algorithm to go astray.

        \subsection{Track selection}
        \label{sect:track_seletion}
        
            In \citet{pfeiferTrailCometTail2022}, we introduced a simple but effective criterion---a miss-rate $\Gamma < 30\%$---to separate genuine from ambiguous tracks, and demonstrated potential applications for our tracking algorithm with tentative first results. While we continue to use the miss-rate criterion as a means of pre-selection, we have now substantially refined the analysis that follows it. 
     
            To derive physical properties of ejected particles, such as their sizes and dynamics, it is essential to know the particle-observer distances. \citet[][and references therein]{chesleyTrajectoryEstimationParticles2020}, for example, were able to retrieve particle-observer distances and reconstruct the full 3D trajectories of particles around asteroid \object{101955 Bennu}; but doing the same for particles near an active cometary nucleus is extremely difficult. For one, in addition to a complex gravity field, active comets have unknown gaseous environments, whose influence on the particles is complex \citep{skorovAccelerationCometaryDust2016, skorovDynamicalPropertiesAcceleration2018, reshetnykDynamicsDustParticles2018, ivanovskiDynamicsAsphericalDust2017, ivanovskiDynamicsNonsphericalDust2017, morenoDynamicsIrregularlyShaped2022}. Secondly, if the (decimeter-sized) particles retain large amounts of water ice after ejection \citep{gundlachActivityCometsUnderstanding2020, davidssonAirfallComet67P2021}, they may even be outgassing themselves and produce a measurable rocket force \citep{kelleyDistributionLargeParticles2013, kelleyErratumDistributionLarge2015, agarwalAccelerationIndividualDecimetresized2016, guttlerCharacterizationDustAggregates2017}. Finally, due to the aforementioned residual pointing fluctuation, our positional data is generally not precise enough to account for such effects and reconstruct 3D trajectories. We hence have to work in the 2D image plane and use a different way to approximate the particle-observer distance.
            
            In \citet{pfeiferTrailCometTail2022}, we made the basic assumption that particles, whose tracks started close to a central active area on the surface of 67P, and whose velocity vector pointed in the same direction as the estimated surface normal $\pm 45\degree$, likely originated from that area. This allowed us to use the nucleocentric distance as a proxy for the particle-observer distance. 
            
            \begin{figure*}[!hb]
                \centering
                \includegraphics[width=\linewidth]{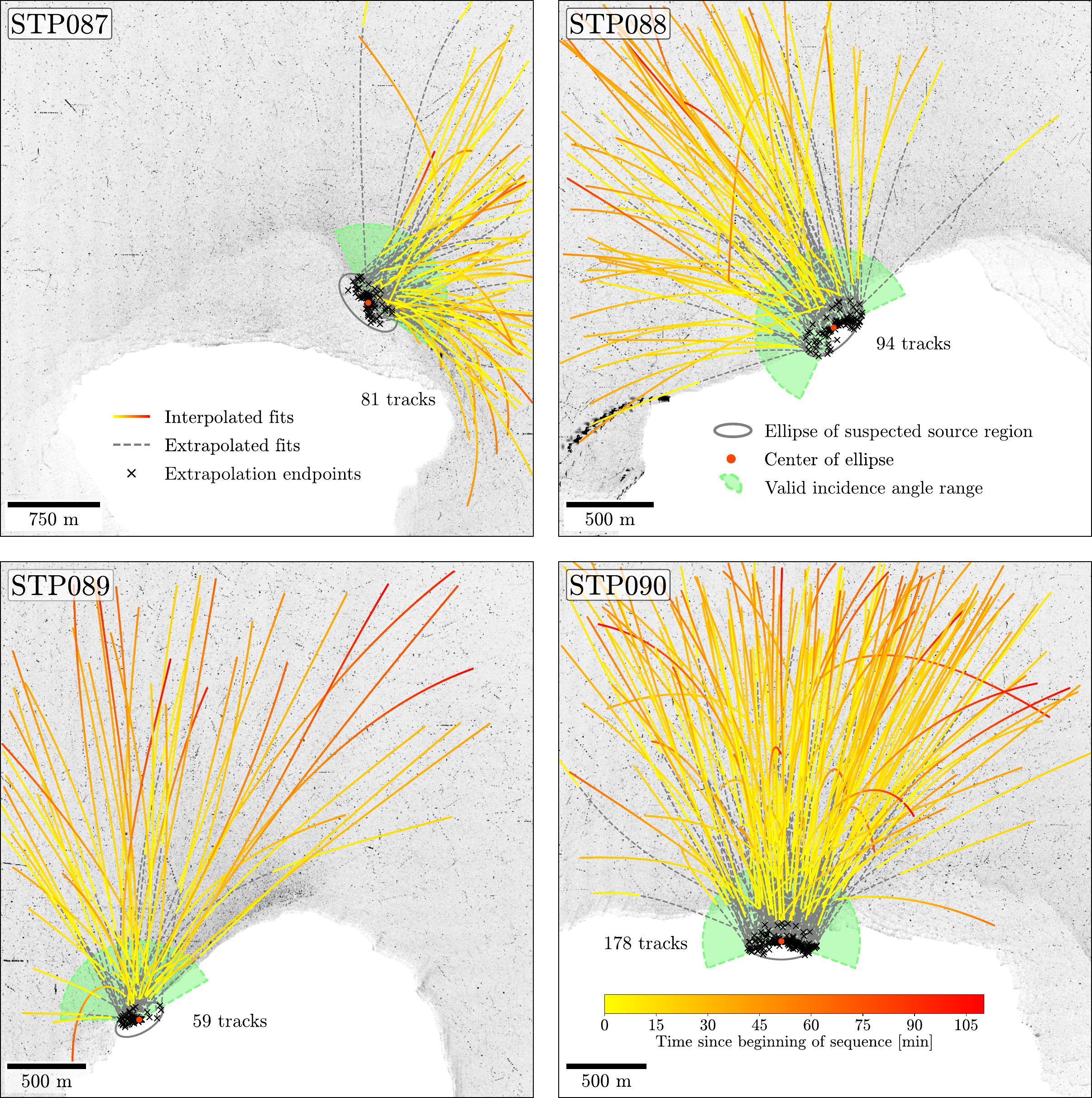}
                \caption{Selected tracks from the four sequences. The fits were extrapolated back in time for a maximum of 30\,min. The extrapolated curves end where they are closest to the ellipse centers. The endpoints are hence not intended to mark the exact ejection times or places.}
                \label{fig:merged_tracking_results}
            \end{figure*}
            
            While we still use the nucleocentric distance as an estimate for the particle-observer distance, we now change how the particles for which we make this assumption are selected. First, we find a concentrated group of particles in the master image. Such groups are key, because their tracks intersect the nucleus within a relatively well-defined area. This allows us to assume that the particles (a) were ejected only recently, and (b) share a common origin (two assumptions that are unreasonable to make about the more randomly or homogeneously distributed tracks). Then, we define an ellipse that roughly outlines their suspected source region in the first image of a sequence, placed on the nucleus close to its limb and slightly following its contour (see Fig.~\ref{fig:sample_images_stp087_stp088}). Next, we extrapolate the 2D fits of the particle tracks back in time for a maximum of 30 minutes. Within this period, the nucleus rotates at most 14.5$\degree$, which results in a displacement of the suspected source region that we still deem acceptable. Lastly, we check whether the tracks intersect with the ellipse within a certain incidence angle range. For those that do, we assume that they likely originated from this area (see Fig.~\ref{fig:merged_tracking_results}). Even though this approach is still relatively simple and some of the selected particles likely originated from somewhere else, in Section~\ref{sect:results_and_discussion}, we show that the general areas are nevertheless plausible.
            
            In the past, it was also suggested that the crossing of particle \enquote{streams} can lead to apparent (artificial) features in the particle number/coma density and the average particle motion \citep[and specifically so in the case of coma simulations using fluid dynamics, e.g.,][]{crifoNearNucleusComaFormed1995, crifoDirectMonteCarlo2005, rodionovAdvancedPhysicalModel2002, zakharovMonteCarloMultifluidModelling2009}. More recently, \citet{shiComaMorphologyComet2018} indeed confirmed that collimated gas and dust flows can result from topographic focusing, but additionally note that apparent jet-like features can also be optical illusions that emerge from projections along the viewing direction. Since we are tracking individual particles however, and thus are not concerned with average coma densities or particle motions within certain volumes along the line of sight, the crossing of particle trajectories is irrelevant in this regard.
            
            During the tracking procedure itself, crossing trajectories are also generally unproblematic. For the tracking algorithm to get confused and jump from one trajectory to the other, they also need to cross at the same time, and have relatively similar speeds, accelerations, and directions. While this can still happen occasionally, such cases are effectively filtered out by our acceptance thresholds and the miss-rate criterion \citep[see][]{pfeiferTrailCometTail2022}. In total, we recovered 11\,858 potential particle tracks from the four image sequences, of which 3626 have miss-rates less than $30\%$. Of those, we traced back 409 to the suspected source regions and visually inspected them several times. We are hence confident that none of the 409 particle tracks that we analyze in the following consist, or were affected by, crossing trajectories. During the entire process, we also did not see any particles colliding or breaking apart.

    \section{Results and discussion} 
    \label{sect:results_and_discussion}

        \subsection{Suspected source regions}
        \label{sect:suspected_source_regions}
        
            \ctable[
                caption = Summarized findings regarding the four suspected source regions.,
                label   = tab:source_regions,
                pos      = b,
                width   = \textwidth,
                notespar,
                star
            ]{lrrrrrrr}{
                \tnote[a]{Number of associated tracks.}
                \tnote[b]{Longitude center.}
                \tnote[c]{Observational period in local time at source region center. Time shift provides local times at source region edges. Universal duration: 03:39.}
                \tnote[d]{Covered surface regions and subregions as defined by \citet{thomasRegionalUnitDefinition2018}.}
                \tnote[e]{Terrain features.}
                \tnote[f]{Differential power-law index of associated particle SFD.}
            }{                \FL  
                Sequence        & $N_\text{t}$\tmark[a]  & $\lambda_\text{c}$\tmark[b]  [$\degree$] & $\tau$\tmark[c] (LT) & Covered surface regions\tmark[d]   & Source characteristics\tmark[e]  & $b$\tmark[f] \ML % \resizebox{!}{9pt}{\Cartouche{\pmglyph{d-w-k-k}}}
                STP087 & 81   & 206   & 13:57--17:35 $\pm$ 2:14   & Im[b], Kn[a, b, c, d], Am[a]   & Rough, boulders, bright spots, jets & $3.8 \pm 0.4$    \ML
                STP088 & 94   & 226   & 11:06--14:45 $\pm$ 1:24    & Kn[d], Am[a]   & Rough, terrace, scarp, jets  & $3.6 \pm 0.4$   \ML
                STP089 & 59   & 211   & 9:54--13:33 $\pm$ 2:00    & Kn[a, d]   & Rough, boulders, bright spots, jets & $3.7 \pm 0.6$      \ML
                STP090 & 178  & 238   & 9:36--13:15 $\pm$ 2:18    & Am[a], Ab, Se   & Smooth, scarp, fractures & $3.4 \pm 0.3$    \LL
            }
            
\begin{figure}[!ht]
                \centering
                \includegraphics[width=\linewidth]{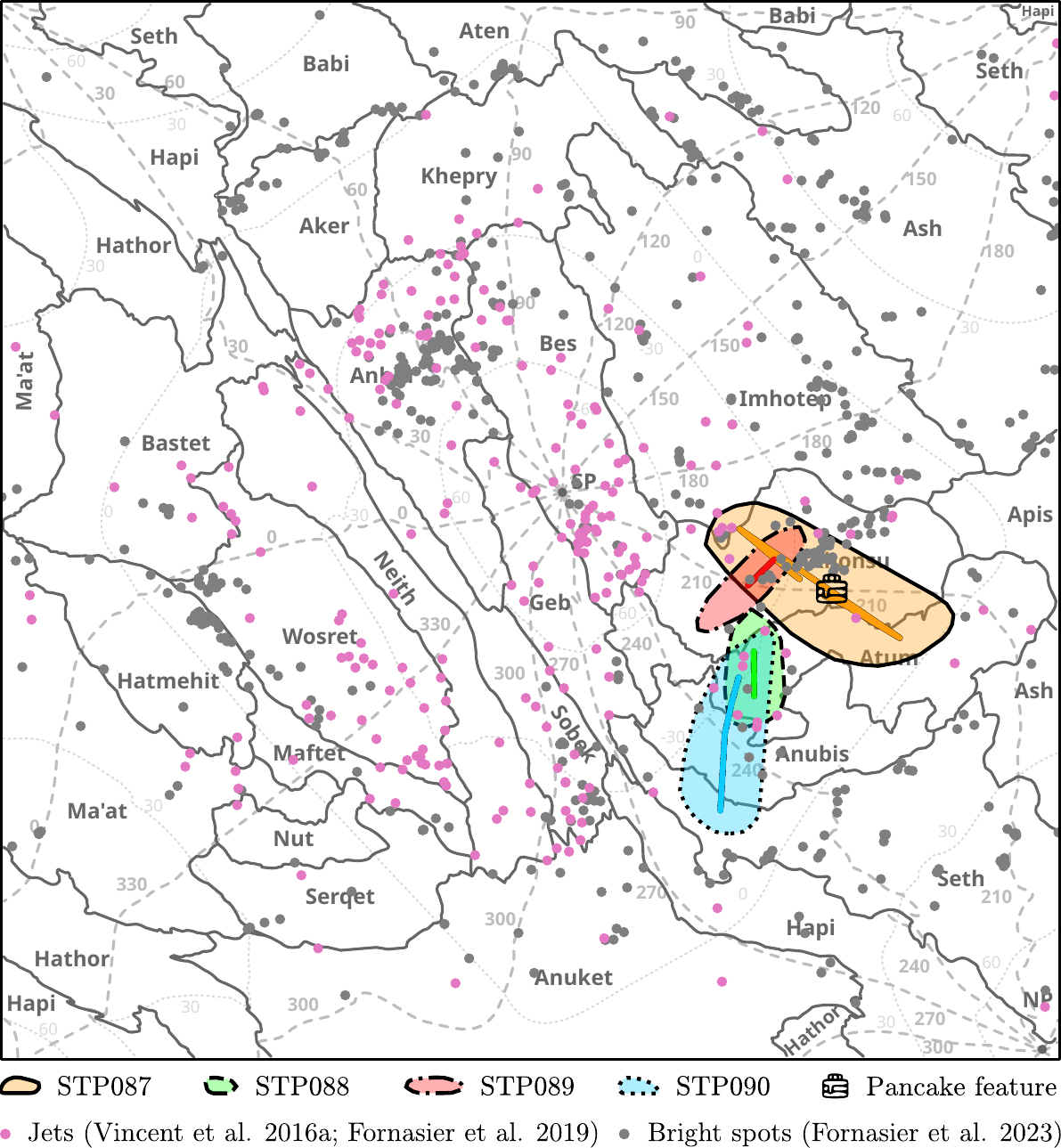}
                \caption[South-centered QuACK-map of 67P's nucleus, centered on its south-pole (SP). The colored shapes show how the ellipses that mark the suspected source regions project onto the surface. The lines in the respective colors indicate the projected ellipse centers. For the geographic data such as surface regions (black lines), latitudes (dotted, light-gray), and longitudes (dashed, gray), we used and slightly modified the publicly available maps provided by \citet{leon-dasiMappingDuckGeological2021}. For the data of the bright spots see also \citet{oklayLongtermSurvivalSurface2017, deshapriyaExposedBrightFeatures2018, hasselmannPronouncedMorphologicalChanges2019, fornasierRosettaComet67P2016, fornasierHighlyActiveAnhur2017, fornasierLinkingSurfaceMorphology2019}.]{QuACK-map of 67P's nucleus, centered on its south pole (SP). The colored shapes show how the ellipses that mark the suspected source regions project onto the surface. The lines in the respective colors indicate the projected ellipse centers. For the geographic data such as surface regions (black lines), latitudes (dotted, light gray lines), and longitudes (dashed, gray lines), the publicly available maps provided by \citet{leon-dasiMappingDuckGeological2021}\footnotemark{} were used (slightly modified). For the data of the bright spots, see also \citet{oklayLongtermSurvivalSurface2017, deshapriyaExposedBrightFeatures2018, hasselmannPronouncedMorphologicalChanges2019, fornasierRosettaComet67P2016, fornasierHighlyActiveAnhur2017}.}
                \label{fig:quack_map}
            \end{figure}

            To get a better idea about where the suspected source regions are located on the nucleus, we used the SPICE toolkit \citep[spacecraft--planet--instrument--c-matrix--events,][]{actonAncillaryDataServices1996, esaspiceserviceRosettaOperationalSPICE2020} to project the ellipses onto the nucleus surface as it was oriented in the first image of each sequence. For every pixel inside the ellipses, we calculated the points of intersection between the corresponding line of sight and the nucleus shape model SHAP5 \citep[][in the \enquote{cheops} reference frame]{jordaGlobalShapeDensity2016}. Due to 67P's concave shape however, not every point on its highly irregular nucleus is uniquely identified by the typically-used equidistant cylindrical map projection. Hence, we instead use the recently developed quincuncial adaptive closed Kohonen (QuACK) projection \citep{griegerQuincuncialAdaptiveClosed2019, leon-dasiMappingDuckGeological2021} to present our data. Figure~\ref{fig:quack_map}, for example, shows how the four suspected source regions project onto the south-centered QuACK map.
            
            As already indicated in Section~\ref{sect:data_selection}, we chose the suspected source regions according to which parts in the master images looked to be most active. Because the diffuse coma and jet-like features have mostly been removed in these images, the choices were based on where the largest, concentrated groups of point-source-like particles appeared to originate from. Interestingly, the suspected source regions do not necessarily coincide with the locations of the strongest diffuse features. In other words, it seems that the ejection of large particles (likely of at least a few centimeters) does not always correlate in location, strength, or orientation with that of small (probably subcentimeter) particles (cf. images in Figs.~\ref{fig:sample_images_stp087_stp088} and \ref{fig:sample_images_stp089_stp090}). 
     
            We also observed this seeming anti-correlation in several other sequences not discussed in detail here. In some of those cases, the spacecraft may have been too far away from the nucleus to distinguish even boulders from the diffuse dust features above its surface, and in others, the phenomenon may be explained by the fact that the sequences were recorded during 67P’s perihelion phase, when the comet was most active and dominated by water ice sublimation \citep[e.g.,][]{combiSurfaceDistributionsProduction2020, lauterGasProduction142020}, which, according to current activity models, can only eject fine dust particles ($ \lesssim 1$\,cm) due to its shallow sublimation front, while \ce{CO2} ice sublimation is responsible for the ejection of larger chunks \citep[e.g.,][]{gundlachActivityCometsUnderstanding2020, fulleHowCometsWork2020, wesolowskiSelectedMechanismsMatter2020, ciarnielloMacroMicroStructures2022, davidssonCO2drivenSurfaceChanges2022}. Yet according to recent thermal modeling by Nicholas Attree \citep[priv. com.]{attreeMoreThermophysicalModelling2023}, the fine-dust erosion driven by water ice sublimation and the ejection of large particles driven by \ce{CO2} ice sublimation cannot happen simultaneously at the same location. The observed lack of large particles within the strong, diffuse dust features may therefore be evidence of this activity-based separation. Notably, \citet{kelleyDistributionLargeParticles2013, kelleyErratumDistributionLarge2015} already observed the same phenomenon for comet \object{103P/Hartley 2}, but they suspected dynamical processes like the rotation of the nucleus, solar radiation pressure, or rocket forces from asymmetric outgassing to be responsible. 

            It may also seem as if the nucleus is not homogeneously, but only locally active in this particle size regime. This observation could be biased however due to our selection criteria of the image sequences (cf. Sect.~\ref{sect:data_selection}), as we were specifically looking for sequences with strong, local activity to fulfill our basic assumptions. Nevertheless, our data clearly show that there are times when inhomogeneous activity occurs.
            
            As illustrated in Figure~\ref{fig:quack_map}, all four suspected source regions overlap substantially. While the recorded local time periods of the four sequences are very similar (with the exception of sequence STP087, see Table~\ref{tab:source_regions}), these areas are not the only ones that were illuminated and visible during the observations (see Figs.~\ref{fig:solar_day_illumination}--\ref{fig:preceding_illumination}, especially because of the different viewing geometries). We could thus have picked areas that lie much farther apart, had there been stronger activity elsewhere. This indicates that the observed activity was likely locally confined and reoccurring. 
            
            \footnotetext{Maps available here: \url{https://doi.org/10.5270/esa-kokoti7}}
            
            It also shows that despite our relatively simple selection process, we likely located the general areas of the source regions correctly. This is further supported by the distinct ejection cones formed by the trajectories that taper toward the ellipse centers (see Fig.~\ref{fig:merged_tracking_results}). To some extent, these ejection cones are likely a consequence of our selection criteria (such as the incidence angle range) and the viewing geometries. Yet their shapes, orientations, and opening angles also suggest that they were likely notably affected by the local surface morphology. Since, to our knowledge, this is the first time such ejection cones have been observed or described, there are currently no theories that explain their occurrence. Their appearance is nevertheless reminiscent of meteorite impacts \citep[e.g.,][]{opikResearchesPhysicalTheory1936, arakawaScientificObjectivesSmall2017, arakawaArtificialImpactAsteroid2020, wadaSizeParticlesEjected2021}, volcanic eruptions \citep[e.g.,][]{tsunematsuEstimationBallisticBlock2016, cigalaDynamicsVolcanicJets2017, cigalaLinkingGasParticle2021, schmidReleaseCharacteristicsOverpressurised2020}, cryovolcanism \citep[e.g.,][]{quickConstraintsDetectionCryovolcanic2013, liuDynamicsDistributionJovian2016, fagentsCryovolcanism2022}, and, of course, cometary jets/geysers/outbursts \citep[e.g.,][]{kellerCometHalleyNucleus1987, wallisTritonEruptionsAnalogous1992, yelleFormationJetsComet2004, ahearnEPOXICometHartley2011, linInvestigatingPhysicalProperties2017, shiClusteredEjectionMetresized2019, wesolowskiSelectedMechanismsMatter2020}. The relatively slow and decentralized release of dust particles however (i.e., resulting not from a singular responsible event, but rather many small, independent ones), from a rough, cometary terrain into the vacuum of space, seems to be unique.
     
            According to Figure~\ref{fig:quack_map}, the selected particles should have been predominantly ejected from the Khonsu, Atum, and Anubis regions, and potentially also from Imhotep and Seth \citep[following the regional definitions of][]{el-maarryRegionalSurfaceMorphology2015, el-maarryRegionalSurfaceMorphology2016, el-maarryRegionalSurfaceMorphology2017, thomasRegionalUnitDefinition2018}. The terrains of these regions are very diverse. Generally, the relatively compact area that is covered by the suspected source regions contains almost every kind of (geomorphological) feature that was observed on the nucleus, be it smooth, or consolidated and outcropped. This includes ridges, terraces, scarps, niches, (coarse or fine grained) dust-blankets, boulders \citep{el-maarryRegionalSurfaceMorphology2015, el-maarryRegionalSurfaceMorphology2016, ferrariBigLobe67P2018, thomasMorphologicalDiversityComet2015, thomasRegionalUnitDefinition2018, leon-dasiMappingDuckGeological2021}, (steep) cliffs \citep{attreeTensileStrength67P2018}, fractures and other linear features \citep{giacominiGeologicMappingComet2016, leeGeomorphologicalMappingComet2016}, bright spots \citep[associated with freshly exposed volatiles,][]{deshapriyaSpectrophotometryKhonsuRegion2016, deshapriyaExposedBrightFeatures2018, fornasierVolatileExposures67P2023}, jet sources \citep{vincentSummerFireworksComet2016, fornasierLinkingSurfaceMorphology2019, laiSeasonalVariationsSource2019}, and potentially even clods or \enquote{goosebumps} \citep{sierksNucleusStructureActivity2015, davidssonPrimordialNucleusComet2016}. Khonsu in particular also contains the striking landmark dubbed \enquote{pancake feature} \citep{el-maarryRegionalSurfaceMorphology2016}, and has been found to be one of 67P's most active regions \citep{deshapriyaSpectrophotometryKhonsuRegion2016, hasselmannPronouncedMorphologicalChanges2019}. 
            
            Due to this variety in morphology, there are several locations within the suspected source regions that could be the origin of the observed activity. In the area near the pancake feature, for example, \citet{vincentSummerFireworksComet2016} and \citet{fornasierLinkingSurfaceMorphology2019} located several jet sources, \citet{fornasierVolatileExposures67P2023} identified many bright spots (see Fig.~\ref{fig:quack_map}), and \citet{hasselmannPronouncedMorphologicalChanges2019} discovered surface changes such as boulder movements and cavity formations. All these occurrences however essentially happened throughout Rosetta's entire rendezvous phase with 67P, and not specifically during our observational period. Nevertheless, the area near the pancake feature substantially overlaps with the suspected source regions (and even the ellipse centers) of sequences STP087 and STP089, which identifies it as a probable source of the particles associated with these regions.  
            
            The suspected source regions of sequences STP088 and STP090, on the other hand, predominantly coincide with Atum and Anubis, which are two vastly different areas. While Atum is an elevated region that mainly consists of consolidated structures like terraces, ridges, and scarps, Anubis is a flat area, which is mostly covered in smooth material, but that also displays some fractures \citep{el-maarryRegionalSurfaceMorphology2015, el-maarryRegionalSurfaceMorphology2016, thomasRegionalUnitDefinition2018, leon-dasiMappingDuckGeological2021}. Even though these areas have not yet been studied as closely as Khonsu, Figure~\ref{fig:quack_map} shows that especially around the suspected source region of sequence STP088, \citet{vincentSummerFireworksComet2016} and \citet{fornasierLinkingSurfaceMorphology2019} have located several jet sources, which might be related to the activity we observed. Additionally, there is a long terrace with a scarp in this area \citep{leeGeomorphologicalMappingComet2016, leon-dasiMappingDuckGeological2021}, so the activity might have also come from an event similar to the cliff collapse discovered at the Aswan site \citep{pajolaAswanSiteComet2016}. 
            
            \begin{figure}[!t]
                \centering
                \includegraphics[width=\linewidth]{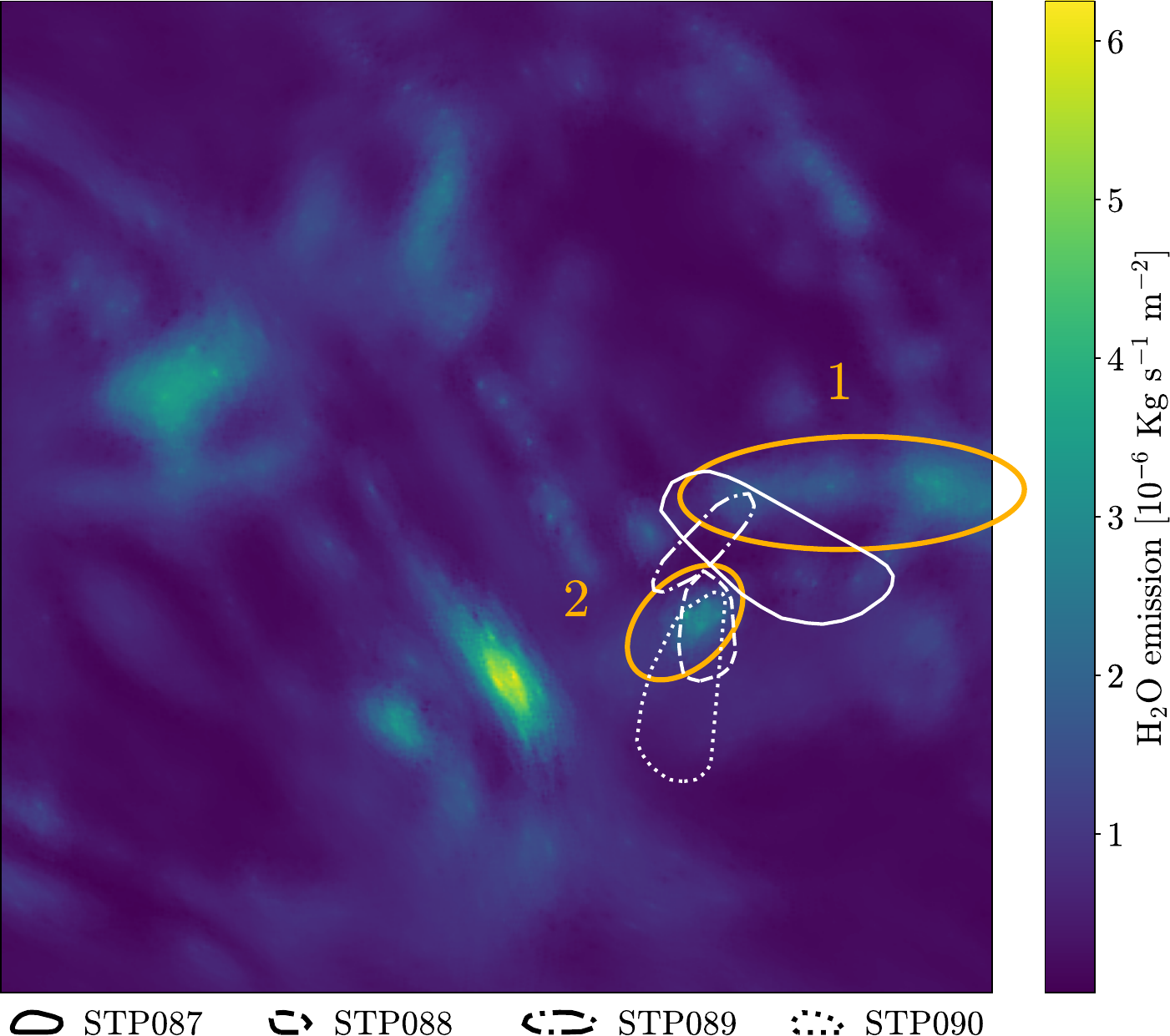}
                \caption{South-centered QuACK-map of the water emission rates predicted by \citet{lauterIceCompositionClose2022}, averaged over the period between December 13, 2015, and January 14, 2016 (data kindly provided by Matthias Läuter and Tobias Kramer). Also shown are the four suspected source regions and the approximate locations of areas 1 and 2 (orange ellipses, drawn in by hand) from \citet{lauterIceCompositionClose2022}.}
                \label{fig:QuACK_emission}
            \end{figure}
            
            In the case of sequence STP090, it is also possible that the activity was caused by retreating scarps of fine material that were reported in this area \citep{el-maarrySurfaceChangesComet2017}, or by the formation and deepening of fractures \citep{leon-dasiMappingDuckGeological2021}, which can facilitate activity \citep{hofnerThermophysicsFracturesComet2017}. STP090's suspected source region also includes steep cliffs located in Seth \citep{attreeTensileStrength67P2018}, but it seems less likely that the observed particles originated from there since the cliffs were not as strongly illuminated during the observational period as the other areas (see Fig.~\ref{fig:solar_day_illumination}).
            
            Finally, the suspected source regions not only lie close to where \citet{combiSurfaceDistributionsProduction2020} estimates the four major volatiles (\ce{H2O}, \ce{CO2}, \ce{CO}, and \ce{O2}) have caused the most erosion between 67P's inbound equinox and August 2016, but they also notably overlap with two areas that \citet{lauterSurfaceLocalizationGas2019, lauterGasProduction142020, lauterIceCompositionClose2022} estimate to be among the most active in water and \ce{CO2} production (and possibly many other species) during the perihelion phase, but also long afterward. Figure~\ref{fig:QuACK_emission} shows the QuACK-map of the water emission rates predicted by \citet{lauterIceCompositionClose2022}. The data was averaged over the period between December 13, 2015, and January 14, 2016, which aligns closely with the overarching time frame of our observations. While the suspected source regions of STP087 and STP089 partly overlap with their area 1, the regions of STP088 and STP090 do so with their area 2.
            
            Overall, we can generally associate the ejected particles with areas that contain rugged terrain or steep slopes, be it scarps, cliffs, or fractures. Such geological features are also considered to favor activity, because they (a) should make it easier for particles to overcome the local gravity and tensile strength that keep them in place \citep{groussinGravitationalSlopesGeomorphology2015, attreeTensileStrength67P2018, attreeThermalFracturingComets2018}, (b) can absorb and store energy more efficiently than flat or smooth areas \citep{hofnerThermophysicsFracturesComet2017}, and (c) do not get covered by loose material that could quench their activity \citep{vincentAreFracturedCliffs2016}. Additionally, we can likely link some of the observed activity to surface areas with a high concentration of jet sources and bright spots, which fits well with our assumptions: Both features indicate (recent) activity, as bright spots may be the remnants of meter-sized, water-ice-enriched blocks \citep[WEBs,][]{ciarnielloMacroMicroStructures2022} that were exposed due to erosion driven by \ce{CO2} ice sublimation.

        \subsection{Particle size-frequency distribution}
        \label{sect:particle_size-frequency_distribution}
        
            Because we associated the particle groups with regions on the nucleus surface, we can use the average nucleocentric distances of the spacecraft (see Table~\ref{tab:meta_data}) as rough estimates for the particle-observer distances. By further assuming that the particles are spherical, we can approximate their equivalent radii via
            \begin{align}
                \label{eq:radius} r = \sqrt{J \frac{r_\text{h}^2 \Delta^2}{R I_\odot}},
            \end{align}
            where $r$ is the particle radius in m, $J$ the particle flux averaged over all its detections in W\,m$^{-2}$\,nm$^{-1}$, $r_\text{h}$ the dimensionless heliocentric distance measured in units of AU, $\Delta$ the particle-observer distance in m, \mbox{$R=0.0021$} the particle reflectance \citep[computed for decimeter-sized particles at a $90\degree$ phase angle using the model from][]{markkanenInterpretationPhaseFunctions2018}, and \mbox{$I_\odot=1.565$\,W\,m$^{-2}$\,nm$^{-1}$} the solar flux in the NAC F22 filter at 1\,AU. 

            For distinct size ranges, it is generally assumed that the size-frequency distribution (SFD) of cometary particles follows a power law. When it comes to particles that are found on the nucleus surface (other than fall-back material), there is good reason to believe that they mostly formed under fractal fragmentation processes like thermal fatigue or gravitational collapse \citep{pajolaSizefrequencyDistributionBoulders2015, pajolaPristineInteriorComet2017, attreeThermalFracturingComets2018, cambianicaQuantitativeAnalysisIsolated2019}, which have been studied extensively for terrestrial material \citep[e.g.,][]{mandelbrotFractalGeometryNature1982, turcotteFractalsFragmentation1986, sanchidrianSizeDistributionFunctions2014, fowlerTheoreticalExplanationGrain2016}. Impact and collisional fragmentation, as seen in asteroids for example, also produce fractal SFDs \citep{dohnanyiCollisionalModelAsteroids1969, hartmannTerrestrialLunarInterplanetary1969, brownTheorySequentialFragmentation1989}, but likely only play a minor role in the case of 67P \citep[so far, evidence for just a single impact  crater has been found,][]{thomasMorphologicalDiversityComet2015}. 
            
            Numerous studies have determined SFDs of 67P's surface material and fitted them with power laws \citep[e.g.,][]{augerGeomorphologyImhotepRegion2015,  mottolaStructureRegolith67P2015, laforgiaGeomorphologySpectrophotometryPhilae2015, pommerolOSIRISObservationsMetersized2015, vincentLargeHeterogeneitiesComet2015, vincentAreFracturedCliffs2016, pajolaSizefrequencyDistributionBoulders2015, pajolaSouthernHemisphere67P2016, pajolaAgilkiaBouldersPebbles2016, pajolaAswanSiteComet2016, pajolaPristineInteriorComet2017, pajolaPebblesBouldersSize2017, pajolaMultidisciplinaryAnalysisHapi2019, deshapriyaSpectrophotometryKhonsuRegion2016, pouletOriginLocalStructures2016,  oklayComparativeStudyWater2016, oklayLongtermSurvivalSurface2017, lucchettiCharacterizationAbydosRegion2016,  lucchettiGeomorphologicalSpectrophotometricAnalysis2017, tangBoulderSizeFrequency2017, hasselmannPronouncedMorphologicalChanges2019, cambianicaQuantitativeAnalysisIsolated2019}. Likewise, power laws have been used for a long time to fit or model the cometary dust environment \citep[e.g.,][]{finsonTheoryDustComets1968, finsonTheoryDustComets1968a, fulleEvaluationCometaryDust1989, clarkReleaseFragmentationAggregates2004, ishiguroCometaryDustTrail2008, kelleyDustTrailComet2008, kelleySPITZEROBSERVATIONSCOMET2009, agarwalDustTrailComet2010, agarwalEvidenceSubsurfaceEnergy2017,  fulleComet67PChuryumovGerasimenko2010, sojaCharacteristicsDustTrail2015, laiGasOutflowDust2016, marschallModellingObservationsInner2016, marschallCliffsPlainsCan2017, marschallComparisonMultipleRosetta2019, marschallDusttoGasRatioSize2020, markkanenInterpretationPhaseFunctions2018}, and more recently to describe the SFDs of particles observed by Deep Impact/EPOXI's camera system around comet \object{103P/Hartley 2} \citep{kelleyDistributionLargeParticles2013, kelleyErratumDistributionLarge2015}, and Rosetta's dust detectors and camera system around comet 67P \citep[e.g.,][]{rotundiDustMeasurementsComa2015, merouaneDustParticleFlux2016, merouaneEvolutionPhysicalProperties2017, hilchenbachCOMET67PCHURYUMOV2016,  agarwalAccelerationIndividualDecimetresized2016, agarwalEvidenceSubsurfaceEnergy2017, fulleEvolutionDustSize2016, rinaldiCometaryComaDust2017}. There is however no proven link between the assumption of a power law and the physical origin of the particle SFD (Marco Fulle, priv. com.). Instead, the power law is mainly of practical use when fitted to a limited size range, because it allows to easily compare different size regimes, and immediately indicates which size range dominates the cross section and volume (mass) distribution \citep[e.g.,][]{fulleComet67PChuryumovGerasimenko2010, fulleEvolutionDustSize2016, rotundiDustMeasurementsComa2015, blumEvidenceFormationComet2017}. We therefore also use power laws to describe the SFDs of our particles.
            
            Generally, if an SFD follows a power law, it means that the radii were drawn from a probability distribution in the form of
            \begin{align}
                \label{eq:powerlaw} p(r) \propto r^{-b},
            \end{align}
            where in our context, the exponent $b$ is often called the (differential) power-law index. A common method to fit a power law to data is to construct a histogram from the data and plot it on a log-log scale with logarithmic bins and counts. In this representation, the data form a straight line with slope $-b$ if the histogram follows a power law, which makes it easy to determine the power-law index via least-squares linear regression, but due to systematic errors (like the arbitrary binning intervals), this approach should generally not be trusted \citep{clausetPowerLawDistributionsEmpirical2009}. 
            
            So instead, we rely on maximum-likelihood estimators (MLEs) and Kolmogorov–Smirnov (KS) tests (see, e.g., \citealp{fellerKolmogorovSmirnovLimitTheorems1948} for KS tests, but also \citealp{snodgrassSizeDistributionJupiter2011}, who nevertheless favor least-square regression). To fit our data, we use the Python-package powerlaw \citep{alstottPowerlawPythonPackage2014}, which is based on the statistical methods of \citet{klausStatisticalAnalysesSupport2011} and \citet[][see also \citealp{virkarPowerlawDistributionsBinned2014}]{clausetPowerLawDistributionsEmpirical2009}. Since our data are drawn from a continuous distribution, Equation~\ref{eq:powerlaw} can be written as
            \begin{align}
                \label{eq:continuous_powerlaw} p(r) = C r^{-b},
            \end{align}
            with normalization constant
            \begin{align}
                \label{eq:normalization_constant} C = (b - 1) r^{b - 1}_\text{min},
            \end{align}
            where $r_\text{min}$ is the lower fitting bound, which is required because $p(r)$ diverges for $r \rightarrow 0$. The MLE of $b$ can then be computed as \citep{muniruzzamanMeasuresLocationDispersion1957}
            \begin{align}
                \label{eq:mle} \hat{b} = 1 + n \left[\sum^n_{i=1} \ln \frac{r_i}{r_\text{min}}\right]^{-1},
            \end{align}
            where $r_i \ge r_\text{min}$ ($i = 1,...,n$) are the measured radii. We additionally used KS tests to find the $r_\text{min}$ and $\hat{b}$ values that produce the best fits (although in the case of sequence STP089, we slightly restricted the allowed range for $r_\text{min}$ due to the small number of data points).\footnote{Strictly speaking, a good fit alone is also not sufficient to conclude that the underlying distribution is truly a power law. In addition to a good theoretical model that explains the data, \citet{clausetPowerLawDistributionsEmpirical2009} recommend elaborate hypothesis testing \citep[see, e.g.,][]{kelleyDistributionLargeParticles2013, kelleyErratumDistributionLarge2015}, but that is outside of the scope of this work.}
    
            \begin{figure}[!ht]
                \centering
                \includegraphics[width=\linewidth]{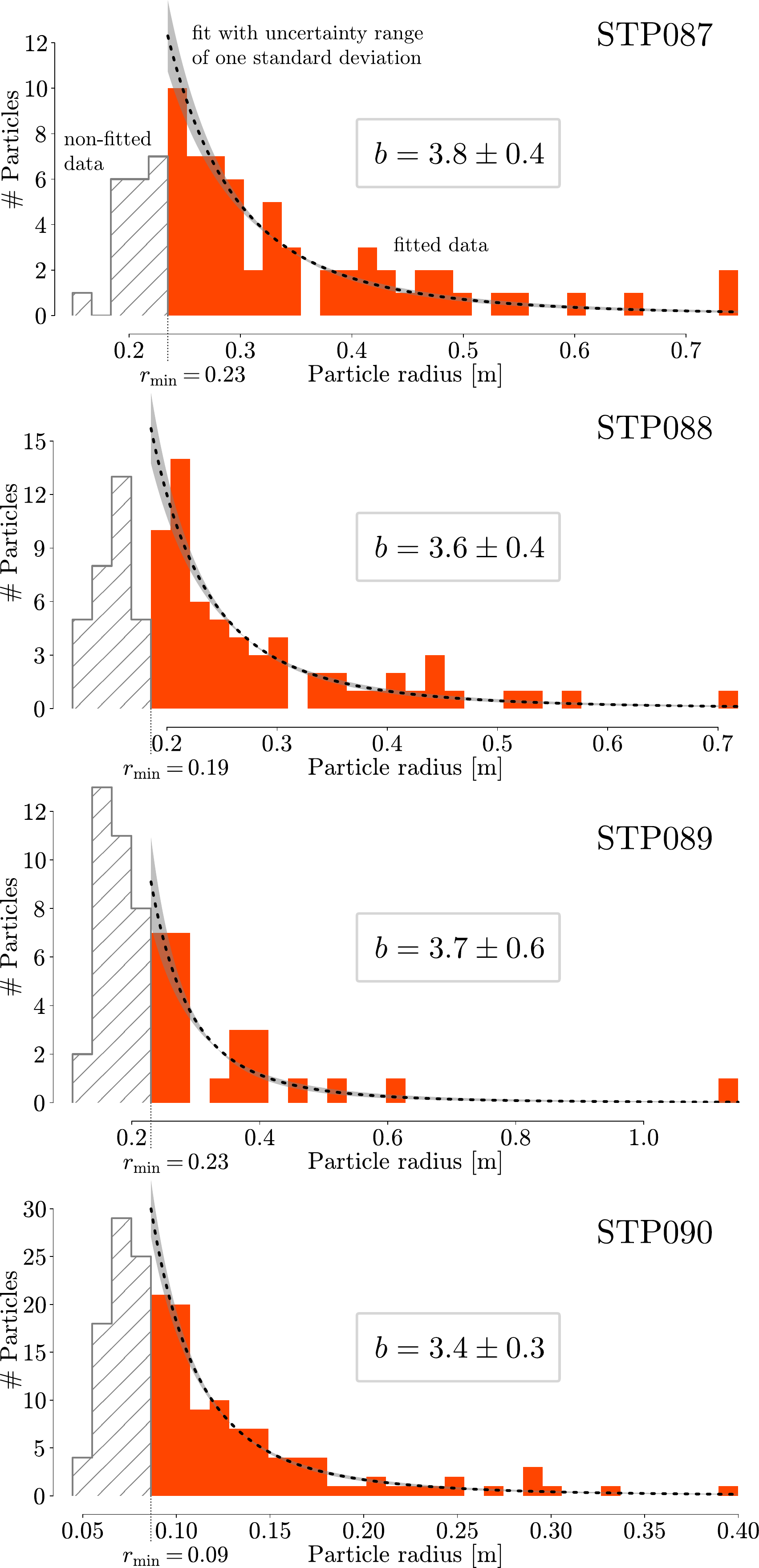}
                \caption{Particle SFDs and fitted power laws of the four particle groups (Fig.~\ref{fig:merged_tracking_results}).}
                \label{fig:radii}
            \end{figure}
            
            Figure~\ref{fig:radii} shows the resulting particle SFDs and fitted power laws. The size ranges of the four SFDs largely overlap, covering mostly the decimeter-scale. Only sequence STP090 differs notably from the other three, with its SFD shifted toward smaller radii, peaking below 10\,cm, and containing the smallest detected particles at roughly 5\,cm. In sequence STP089 on the other hand, we detected the largest particle with an equivalent radius of about 1.15\,m. Many particles strongly fluctuate in brightness over the course of their tracks however (in large part likely because they are nonspherical rotators), so their mean-flux-based radii only represent the most likely average values. In contrast, the errors from residual background signals and inaccurate particle-observer distances, as well as the uncertainty in the albedo, are typically much smaller than these fluctuations. Our results therefore remain sufficiently precise, and following the terminology of \citet{pajolaAgilkiaBouldersPebbles2016}, the particles classify as either pebbles ($<25$\,cm) or boulders ($>25$\,cm). 
            
            The values of the fitted power-law indices range from $3.4 \pm 0.3$ to $3.8 \pm 0.4$ and are all within each others' error bars, suggesting that most of the mass is contained in the larger particles. Yet most of the large particles were likely not fast enough to leave the gravitational field of the nucleus (see Figs.~\ref{fig:speeds_vs_radii} and \ref{fig:accelerations_vs_radii} in Sect.~\ref{sect:particle_dynamics}), and should have later fallen back onto its surface. Farther out in the coma, the SFDs in this size range might hence be a little steeper. So far however, there are no other studies that report power-law indices for coma particle SFDs in a similar size range (or time period) to corroborate our findings. 
            
            Nevertheless, our index values agree notably well with those obtained for submillimeter-sized particles, despite the likely fundamentally different ejection mechanisms: \citet{fulleEvolutionDustSize2016} report $b \approx 3.7$ for dust particles captured during the perihelion phase, and \citet{merouaneDustParticleFlux2016} report $b = 3.1 \pm 0.5$ for those captured between perihelion and April 2016 (these are currently the two observational periods that most closely match ours and for which $b$-values exist). 
            
            Regarding surface material, the most relevant $b$-values were obtained by \citet{deshapriyaSpectrophotometryKhonsuRegion2016} and \citet{hasselmannPronouncedMorphologicalChanges2019}, who counted boulders in the Khonsu region (no studies exist yet for Atum or Anubis). \citet{deshapriyaSpectrophotometryKhonsuRegion2016} surveyed the entire region using images recorded pre-perihelion on May 2, 2015, and report $b = 4.1 +\!0.2/\!-\!0.3$ for boulders ranging from 7 to 40\,m, whereas \citet{hasselmannPronouncedMorphologicalChanges2019} studied a small patch near the pancake feature in images recorded post-perihelion on June 25, 2016, and report $b = 2.6 \pm 0.01$ for boulders ranging from 1.2 to 10\,m. This significant flattening of the particle SFD over the course of the perihelion phase might indicate that most of the smaller boulders were removed from the surveyed area, while most of the larger ones remained or fell back onto it. Since our observations were recorded post-perihelion but well before June 25, 2016, the fact that our power-law indices lie in between those reported by \citet{deshapriyaSpectrophotometryKhonsuRegion2016} and \citet{hasselmannPronouncedMorphologicalChanges2019} might be a reflection of this particle SFD transition. 
            
            Overall, our particles may therefore be a link between the debris studied on 67P's surface and the particles observed and collected in its coma. Not only do they cover a size range that on the upper end overlaps with that of the typically much larger boulders counted on the nucleus surface, and on the lower end with that of the typically much smaller particles detected farther out in the coma \citep[see also][]{guttlerCharacterizationDustAggregates2017, ottDustMassDistribution2017, frattinObservationalConstraintsDynamics2021, lemosDistributionDynamicsDecimetresized2023, lemosEjectionDynamicsAggregates2024}; but from an evolutionary standpoint, they also sit between the relatively pristine surface material which has not yet been ejected (excluding fall-back material), and the more processed material found in the coma and beyond which has escaped the nucleus gravity.

        \subsection{Illumination and surface acceleration}
        \label{sect:illumination_and_surface_acceleration} 
        
            Because solar irradiation fundamentally drives the activity of comets, we created QuACK-maps that show 67P's average surface illumination during the time periods when our sequences were recorded, to estimate the insolation that our suspected source regions received. The maps were generated using the same approach as described by \citet{griegerQuincuncialAdaptiveClosed2019}, which accounts for the relative orientation between the solar irradiation vector and the surface normals of the QuACK shape model tiles, as well as self-shadowing effects. Because the QuACK shape model is relatively rough however (160\,000 tiles), the resulting illumination maps are somewhat blurred. In reality, small-scale surface parts, such as cliff faces, may receive insolation that is much stronger than the average \citep{pajolaPristineInteriorComet2017}. Activity from such sites may be significantly enhanced, which our employed model would not reflect, but since we are mainly interested in general areas, the illumination maps are adequate. 
            
\begin{figure}[!ht]
                \centering
                \includegraphics[width=\linewidth]{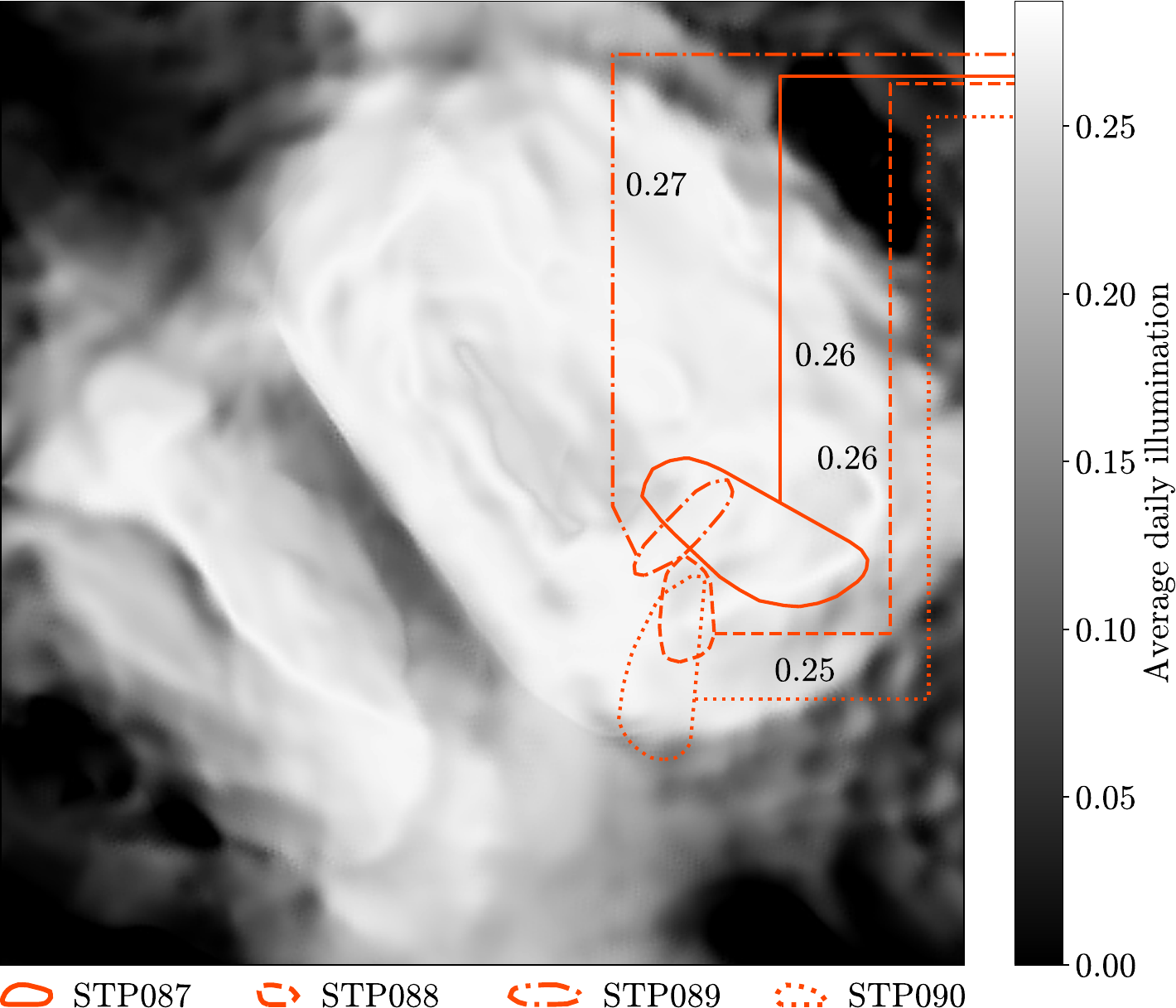}
                \caption{South-centered QuACK-map of the average nucleus illumination received over a whole comet day at the time of sequence STP088 (December 26, 2015). The intensity is measured in units of the average illumination at 1\,AU from the Sun at the equator of the Earth over one Earth-day during equinox. The orange shapes and lines indicate the mean values within the suspected source regions.}
                \label{fig:solar_day_illumination}
            \end{figure}

\begin{figure*}[!ht]
                \centering
                \includegraphics[width=\linewidth]{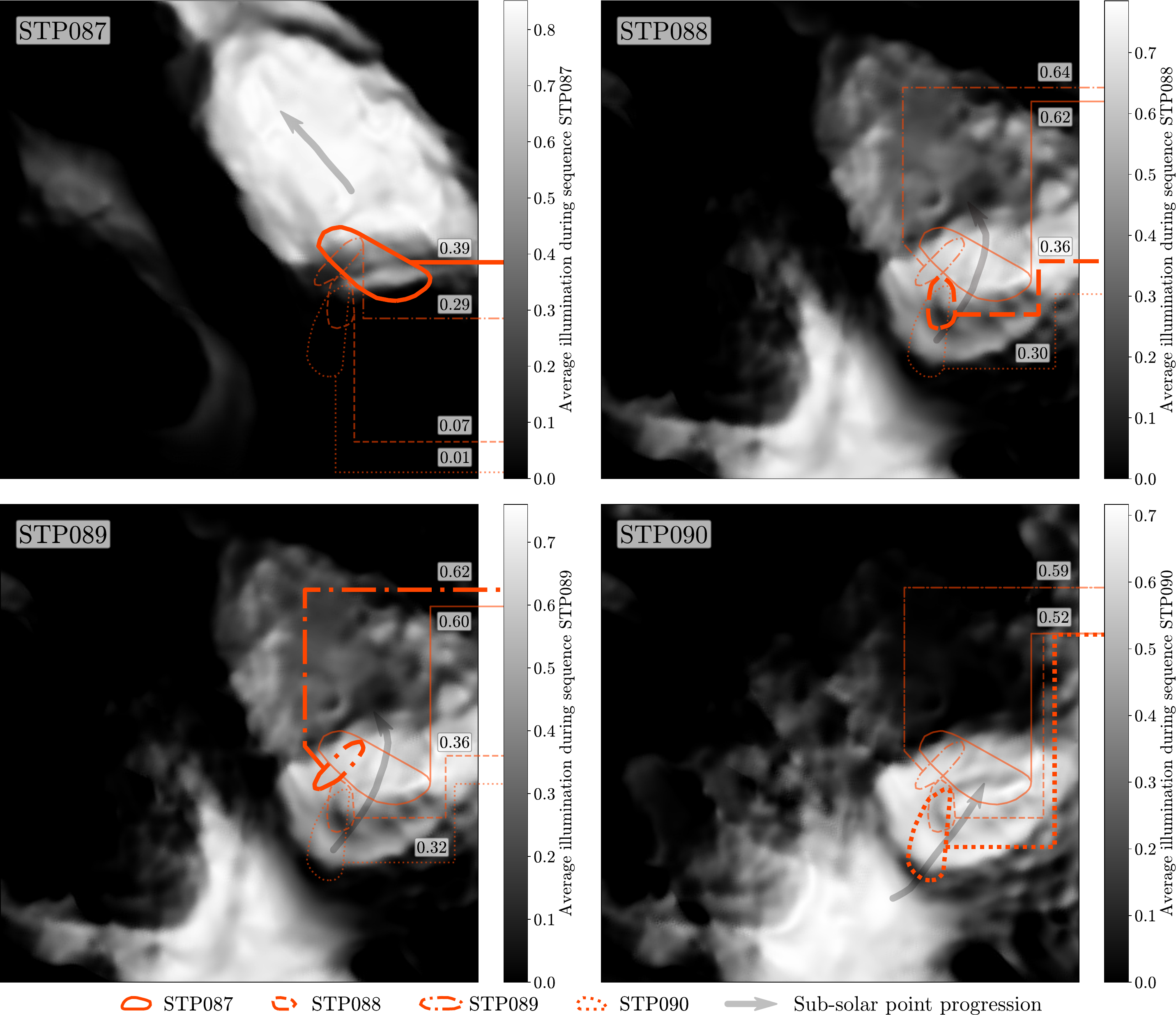}
                \caption{South-centered QuACK-maps of the average surface illumination received during the respective observational periods of the four sequences. The intensity units are the same as in Fig.~\ref{fig:solar_day_illumination}. The mean values within the suspected source regions are indicated by the orange shapes and lines, with the areas that were selected during the respective sequences highlighted in bold. The values here are generally higher than in Fig.~\ref{fig:solar_day_illumination} because here the illumination was averaged over time periods when the areas were mostly in sunlight, while in Fig.~\ref{fig:solar_day_illumination} the illumination was averaged over a whole day-and-night cycle.}
                \label{fig:all_sequence_illumination}
            \end{figure*}

\begin{figure*}[!ht]
                \centering
                \includegraphics[width=\linewidth]{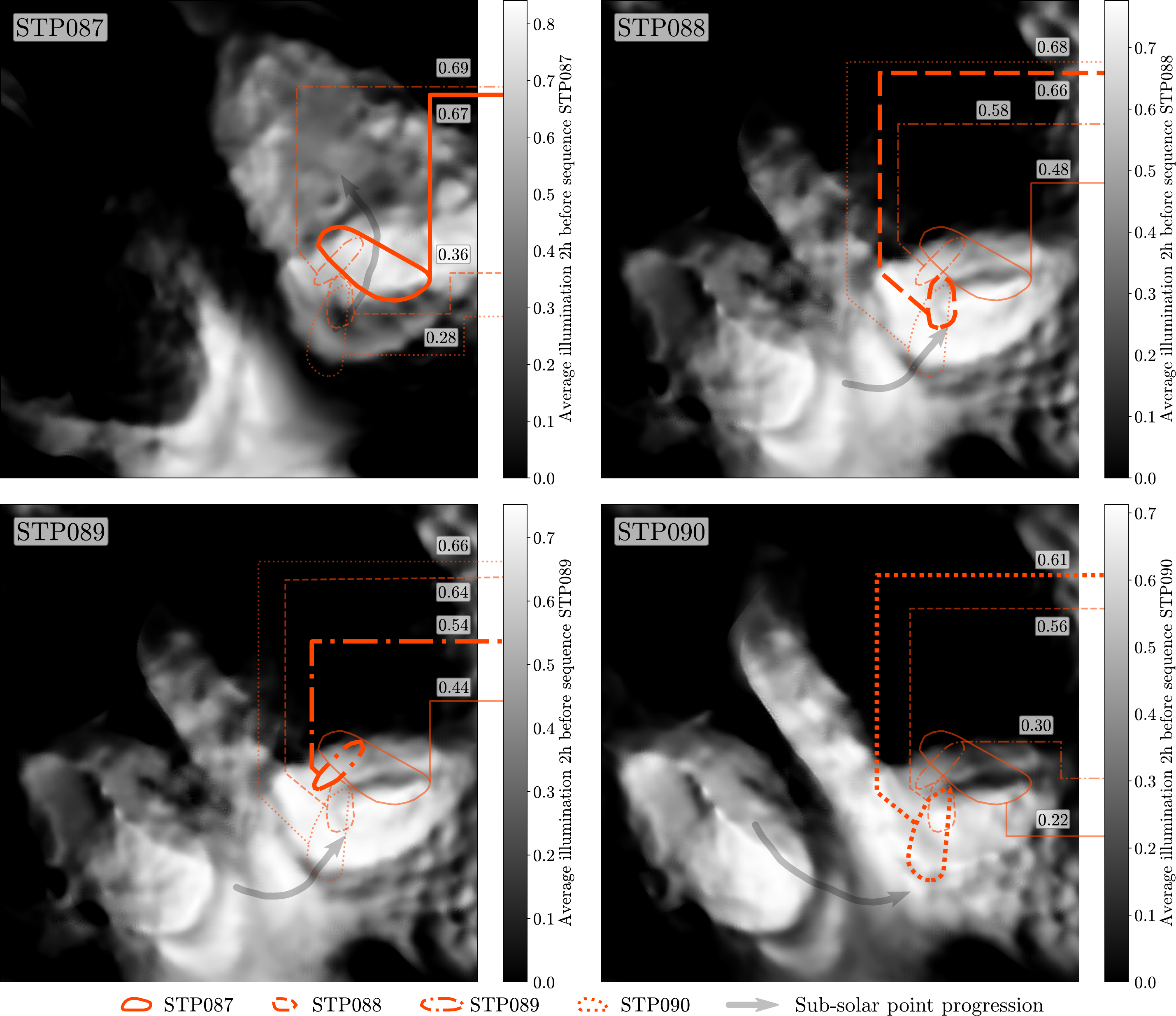}
                \caption{South-centered QuACK-maps of the average surface illumination received during the two hours prior to the respective observational periods. The figure elements and intensity scale are analogous to those in  Fig.~\ref{fig:all_sequence_illumination}.}
                \label{fig:preceding_illumination}
            \end{figure*}

            Because the four sequences were recorded within less than a month (which corresponds to a change in sub-solar latitude of less than six degrees), the illumination conditions averaged over a whole comet day are almost identical on the respective dates and differ mostly in overall intensity. Figure~\ref{fig:solar_day_illumination} shows the average daily illumination for sequence STP088, which is a good representation for the entire period as its sub-solar latitude is at the center of the covered value range (see Table~\ref{tab:meta_data}). Additionally, we created two more illumination maps for each of our sequences. The first set shows the average illumination received solely during the observational periods of the respective sequences (see Fig.~\ref{fig:all_sequence_illumination}); but because our particles were likely ejected up to half an hour before the sequences were recorded, and because the activity does not immediately follow the exposure to sunlight as temperature and gas pressure first need to build up, the other set shows the average illumination received over the two hours (roughly 3\,h and 51\,min in local time) leading up to the observational periods (see Fig.~\ref{fig:preceding_illumination}).
            
            According to Figure~\ref{fig:solar_day_illumination}, the suspected source regions clearly received some of the most sunlight over the course of a full comet day at that time. Conversely, Figure~\ref{fig:all_sequence_illumination} shows that this was not always the case during the observational periods. Surface areas other than the suspected source regions were illuminated just as, or even more strongly during these periods, but did not show nearly the same levels of activity in our image sequences. Yet if the back-extrapolations of our tracks are approximately correct, most of our particles should have been ejected up to half an hour before the observations started (or possibly even earlier). Figure~\ref{fig:preceding_illumination} shows that during the two hours prior to the observations, the illumination conditions for other areas (such as the neck, the small lobe, or Hapi) were a lot different, while the suspected source regions were already well illuminated. This further supports our choice of their general locations. Yet since we did not observe similar activity from areas that received comparable insolation (during either period), it also suggests that strong illumination is necessary, but not sufficient to explain the observed activity. 
    
            \begin{figure*}[!ht]
                \centering
                \includegraphics[width=\linewidth]{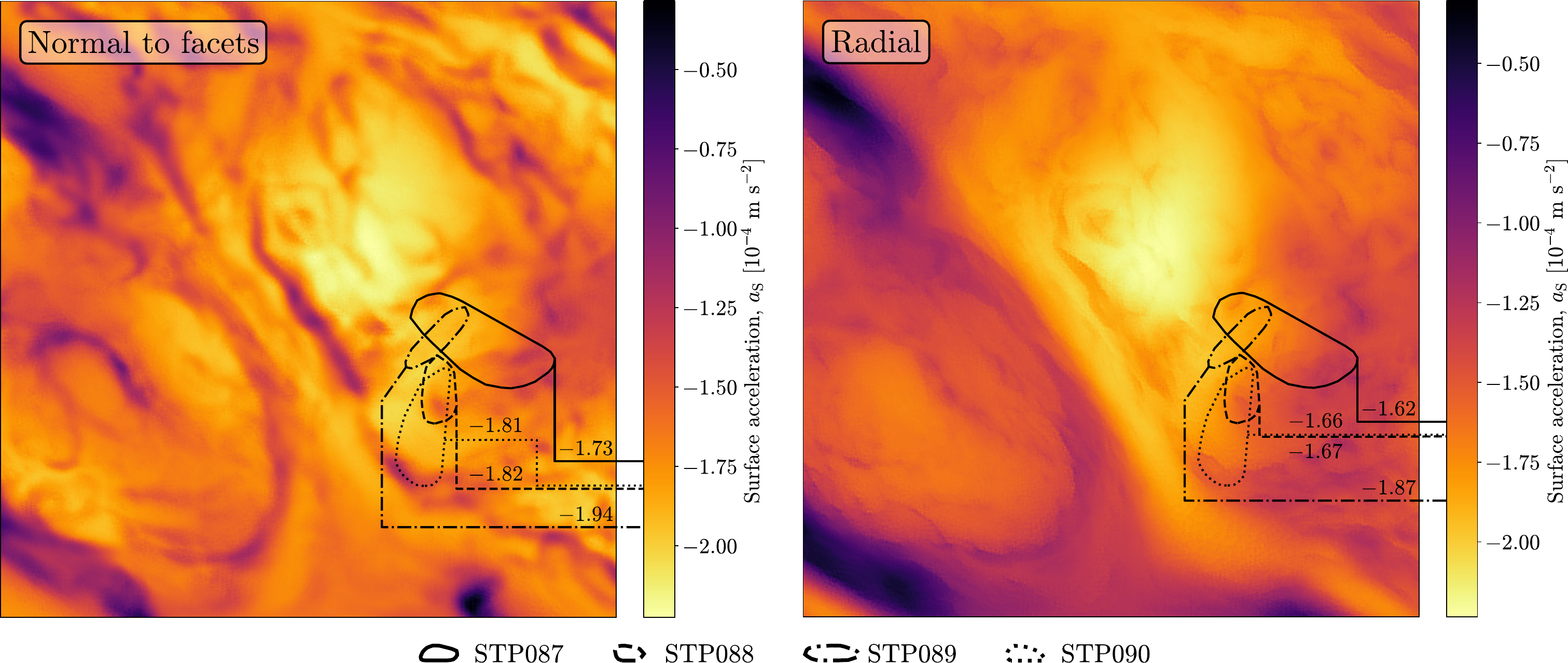}
                \caption{South-centered QuACK-maps of the approximate surface acceleration, $a_\text{S}$, on 67P's nucleus. Component parallel to the facet normal on the left, radial component on the right. The mean values within the suspected source regions are also indicated.}
                \label{fig:surface_acceleration}
            \end{figure*}
            
            We thus also investigated the intensity of the local surface acceleration, $a_\text{S}$ (i.e., the sum of gravity and centrifugal acceleration). Figure~\ref{fig:surface_acceleration} shows two QuACK-maps of $a_\text{S}$, once normal to the shape model facets and once in radial direction. In both cases, the average downward accelerations within the suspected source regions are relatively high, which means that they cannot explain the localized release of decimeter-sized particles either. The local surface composition and structure therefore seem to be the most probable causes. For one, activity-enhancing topographies as discussed in Section~\ref{sect:suspected_source_regions} may play a role \citep[see also][who use models to explore how structural parameters such as porosity and dust layer thickness can affect the gas production rate]{reshetnykTransportCharacteristicsNearSurface2021, reshetnykTransportCharacteristicsHierarchical2022, skorovEffectVaryingPorosity2021, skorovEffectHierarchicalStructure2022, skorovSensitivityModelledCometary2022, skorovCometarySurfaceDust2023}, but more importantly, we regard a local overabundance of \ce{CO2} ice as the most likely driver of decimeter-sized particle ejection \citep[in line with current ejection models, e.g.,][]{gundlachActivityCometsUnderstanding2020, fulleHowCometsWork2020, wesolowskiSelectedMechanismsMatter2020, ciarnielloMacroMicroStructures2022, davidssonCO2drivenSurfaceChanges2022}.

        \subsection{Particle dynamics}
        \label{sect:particle_dynamics}

            From the polynomial fits to the particle tracks we immediately obtain the projected particle velocities and accelerations. Initially, they are measured in terms of the image coordinate system (e.g., in px\,s$^{-1}$), so to translate them into physical units, we again need to know the particle-observer distance and thus again use the nucleocentric distance as a proxy (see also Table~\ref{tab:meta_data}).
            
\begin{figure*}[!ht]
                \centering
                \includegraphics[width=\linewidth]{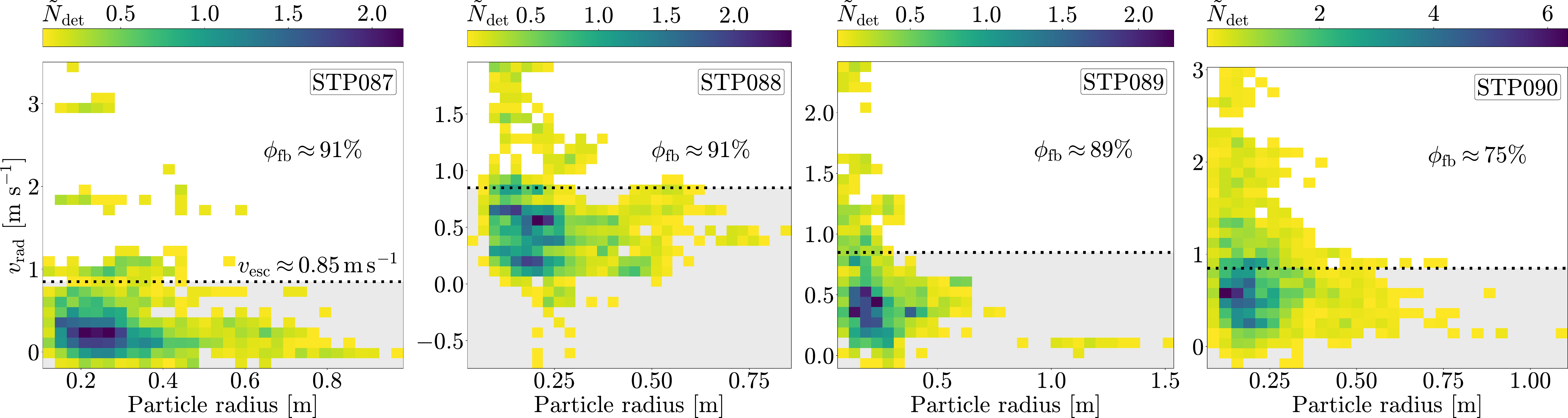}
                \caption{2D histograms of the projected radial particle velocities, $v_\text{rad}$, as a function of particle radius (for the selected particles, see Fig.~\ref{fig:merged_tracking_results}). Because the polynomial fits to the particle tracks also approximate the particle accelerations, instead of using values averaged over a track, the particle velocities and radii were determined for each detection of a track individually and the results were weighted by the number of detections of the respective track (residence time weighting).
                The histogram data thus represent the weighted number of detections, $\tilde{N}_\text{det}$. Also shown are the approximate escape speeds at the suspected source regions, $v_\text{esc}$ (dotted line), and the corresponding maximum fall-back fractions, $\phi_\text{fb}$ (gray area).}
                \label{fig:speeds_vs_radii}
            \end{figure*}

            Figure~\ref{fig:speeds_vs_radii} shows how the radial components (relative to the nucleus' center of mass) of the resulting projected particle velocities, $v_\text{rad}$, distribute as a function of particle radius, and how they compare to the local escape speed, $v_\text{esc}$. Assuming a local surface acceleration at the suspected source regions of approximately $-1.8 \cdot 10^{-4}$\,m\,s$^{-2}$ (based on the data shown in Fig.~\ref{fig:surface_acceleration}), we find that $v_\text{esc} \approx 0.85$\,m\,s$^{-1}$. Overall, roughly $75$ to $91$\% of the selected particles are slower than this escape speed. Some particles even have negative radial velocities, reflecting the fact that some of the tracks in Figure~\ref{fig:merged_tracking_results} already curve back toward the nucleus. Although these velocities are merely projected, the percentages of particles slower than the escape speed directly provide upper-limit estimates of the fall-back fractions, $\phi_\text{fb}$. In particular, almost all particles with $r \gtrsim 40$\,cm seem to be too slow to leave the gravitational field of the nucleus (although  some of them may still reach escape speed later on due to gas drag acceleration). These particles may for example contribute to the dust blankets found on the northern hemisphere of the nucleus such as the Hapi region \citep[e.g.,][]{thomasRedistributionParticlesNucleus2015, kellerInsolationErosionMorphology2015, kellerSeasonalMassTransfer2017, laiGasOutflowDust2016, cambianicaTimeEvolutionDust2020, davidssonAirfallComet67P2021}.
     
            The sizes and radial velocities of the particles that are faster than the escape speed, on the other hand, generally fit well with those of particles detected farther out in the coma by \citet{ottDustMassDistribution2017}. Their particles are mostly centimeter-sized and on average about twice as fast as ours, which seems reasonable given that our particles are mostly decimeter-sized and still accelerating away from the nucleus.
    
\begin{figure*}[!ht]
                \centering
                \includegraphics[width=\linewidth]{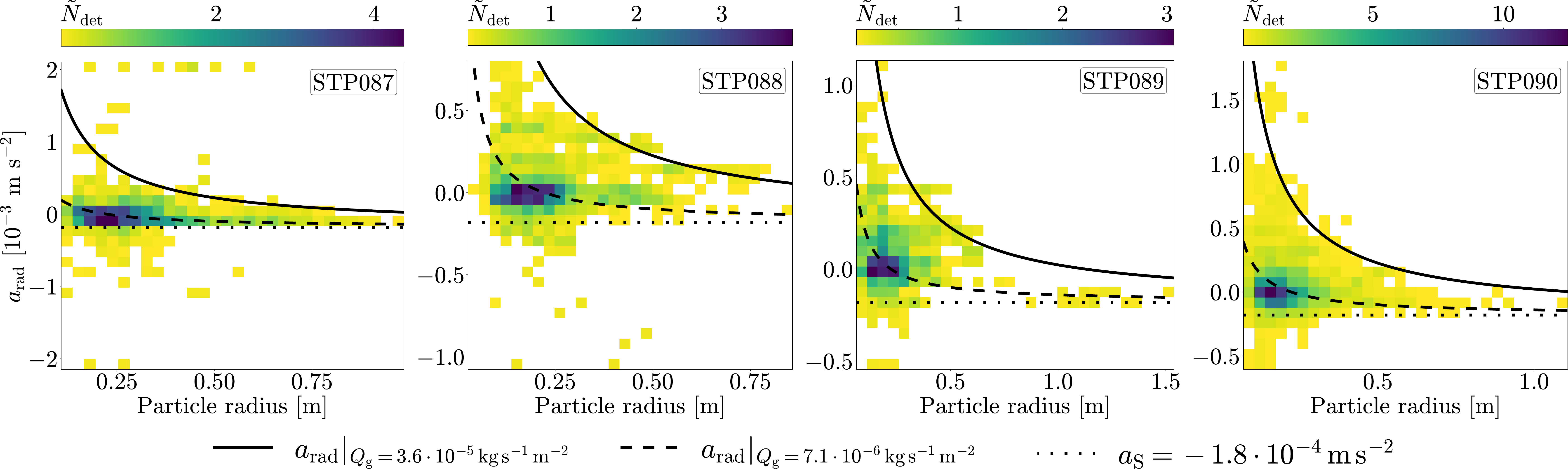}
                \caption{2D histograms of the projected radial particle accelerations, $a_\text{rad}$, as a function of particle radius (for the selected particles, see Fig.~\ref{fig:merged_tracking_results}). The histogram values were determined analogously to the description in Fig.~\ref{fig:speeds_vs_radii}. Also shown are the predicted surface acceleration, $a_\text{s}$ (dotted line, see also Fig.~\ref{fig:surface_acceleration}), and two models (dashed and solid curves) of the particle acceleration, as described by Eqs.~\ref{eq:acceleration2} and \ref{eq:xi_factor}. The parameters for both curves are based on our gas models (see Sect.~\ref{sect:dust_coma_simulations}), but for the solid curve,   a local gas production rate was assumed that is about five times higher.}
                \label{fig:accelerations_vs_radii}
            \end{figure*}

            This is illustrated by Figure~\ref{fig:accelerations_vs_radii}, which shows the projected radial particle accelerations as a function of particle radius. In this context of decimeter-sized particles, we assume that there are mainly three forces acting along the radial direction: gravity, centrifugal force, and gas drag. Solar radiation pressure and solar tides are generally several orders of magnitude weaker than these three, and therefore negligible. Rocket forces exerted by the particles themselves due to asymmetric outgassing, on the other hand, are difficult to estimate \citep[see, e.g.,][]{kelleyDistributionLargeParticles2013, kelleyErratumDistributionLarge2015, agarwalAccelerationIndividualDecimetresized2016, guttlerCharacterizationDustAggregates2017}. The particles' sublimation rate and in particular their ice fraction are still topics of debate, and may be high or insignificant, depending on the model \citep[e.g.,][]{davidssonPrimordialNucleusComet2016, davidssonAirfallComet67P2021, blumEvidenceFormationComet2017, blumFormationComets2022, gundlachActivityCometsUnderstanding2020, fulleHowCometsWork2020, cambianicaTimeEvolutionDust2020, choukrounDusttoGasRefractorytoIceMass2020, ciarnielloMacroMicroStructures2022}. We thus also ignore rocket forces \citep[for discussions of other minor forces that act in such scenarios, see, e.g.,][]{chesleyTrajectoryEstimationParticles2020, jiangMotionDustEjected2020}, and model the radial component of the particle acceleration, $a_\text{rad}$, via
            \begin{align}
                \label{eq:acceleration} a_\text{rad} = \frac{1}{m} \left(F_\text{G} + F_\text{C} + F_\text{D} \right),
            \end{align}
            where $m$ is the particle mass, $F_\text{G}$ and $F_\text{C}$ are the (radial components of) gravitational and centrifugal force, respectively, and
            \begin{align}
                \label{eq:gas_drag} F_\text{D} = \frac{1}{2}C_\text{D} m_\text{g} n_\text{g} \sigma_\text{p} |v_\text{g} - v_\text{p}|(v_\text{g} - v_\text{p})
            \end{align}
            is the (radial component of) gas drag, where $C_\text{D}$ is the (dimensionless) drag coefficient, $m_\text{g}$ the mass of a gas molecule, $n_\text{g}$ the gas number density, $\sigma_\text{p}$ the particle cross section, and $v_\text{g}$ and $v_\text{p}$ are the radial velocities of the gas and the particles, respectively. Plugging Equation~\ref{eq:gas_drag} into Equation~\ref{eq:acceleration} and assuming $v_\text{g} \gg v_\text{p}$, as well as spherical particles with bulk density $\rho_\text{p}$ and radius $r$, we get
            \begin{align}
                \label{eq:acceleration2} a_\text{rad} = a_\text{S} + \xi r^{-1},
            \end{align}
            where $a_\text{S} = (F_\text{G} + F_\text{C})\,m^{-1}$ is the surface acceleration, and roughly equal to $-1.8 \cdot 10^{-4}$\,m\,s$^{-2}$ at our suspected source regions according to the data shown in Figure~\ref{fig:surface_acceleration}, and
            \begin{align}
                \label{eq:xi_factor} \xi = \frac{3}{2} C_\text{D} Q_\text{g} v_\text{g} \rho_\text{p}^{-1},
            \end{align}
            where $Q_\text{g} = \frac{1}{4} m_\text{g} n_\text{g} v_\text{g}$ \citep{birdMolecularGasDynamics1994} is the local gas production rate at the suspected source regions.
            
            The parameters that constitute $\xi$, however, are difficult to constrain. The particle density, for instance, might be significantly lower than the bulk density of the nucleus, if the particles lost most of their volatiles but kept their structure and thus volume intact. Likewise, the drag coefficient also depends on particle shape, macro porosity, and other parameters, and could consequently vary a lot \citep[e.g.,][]{skorovAccelerationCometaryDust2016, skorovDynamicalPropertiesAcceleration2018, ivanovskiDynamicsAsphericalDust2017, ivanovskiDynamicsNonsphericalDust2017, reshetnykDynamicsDustParticles2018}, and in correlation with the gas velocity and number density, the local gas production rate could even differ by one or two orders of magnitude \citep[see also][]{marschallLimitationsDeterminationSurface2020, marschallCometaryComaeSurfaceLinks2020, marschallNeutralGasComa2023}. 
            
            Nevertheless, from fitting Equation~\ref{eq:acceleration2} visually to our data, we find that $\xi \approx 2 \cdot 10^{-4}$\,m$^{2}$\,s$^{-2}$ describes the upper limit of the measured $a_\text{rad}$ rather well for all four sequences (see solid curves in Fig.~\ref{fig:accelerations_vs_radii}). To achieve such a value, we for example propose the following parameter combination: $C_\text{D} = 4$, $Q_\text{g} = 3.6 \cdot 10^{-5}$\,kg\,s$^{-1}$\,m$^{-2}$, $v_\text{g} = 500$\,m\,s$^{-1}$, and $\rho_\text{p} = 533$\,kg\,m$^{-3}$. While the particle density value is a conservative estimate based on the nucleus bulk density \citep{patzoldHomogeneousNucleusComet2016} and may be notably lower, the other parameter values are based on our water activity model \citep[$m_\text{g} = m_\text{H\textsubscript{2}O} = 3 \cdot 10^{-26}$\,kg, $n_\text{g} = 1.9 \cdot 10^{18}$\,m$^{-3}$; see Fig.~\ref{fig:gas_models}, Sect.~\ref{sect:dust_coma_simulations}, and][]{marschallDusttoGasRatioSize2020}, assuming that the local gas production is about five times higher than our model prediction ($Q_\text{g} = 7.1 \cdot 10^{-6}$\,kg\,s$^{-1}$\,m$^{-2}$, which is also very similar to the peak average production rates estimated by \citet{lauterIceCompositionClose2022}, see Fig.~\ref{fig:QuACK_emission} and dashed curves in Fig.~\ref{fig:accelerations_vs_radii}). This fits well with our observations of strong, localized activity, likely driven by \ce{CO2} ice sublimation. Additionally, the lower bound of the measured radial accelerations also agrees particularly well with the estimated surface acceleration. 
    
            There are some measurements in Figure~\ref{fig:accelerations_vs_radii} however, that clearly lie outside these theoretical bounds. More curiously, there is an apparent trend for outliers with radii $\lessapprox 40$\,cm to have radial accelerations that are much lower than the predicted surface acceleration. To investigate this trend, we looked at all the particles that exhibited downward accelerations smaller than $a_\text{S}$ at least once while they were observed. The corresponding tracks indicate that some of the outliers likely belong to particles that do not actually originate from the suspected source regions, but were instead flying in the fore- or background and just passed through the ellipses by chance. Others may have inaccurate back-extrapolated fits, and a third group of particles seem to \enquote{orbit} around the nucleus at a distance and orientation where the \enquote{downward} direction as defined by the ellipses is inadequate, and so they only seem to have such strong downward accelerations due to the 2D projection. These causes should also apply to the other outliers, but the described trend is likely at least in part a statistical effect, since most of our particles lie in this size regime (see also Fig.~\ref{fig:radii}), and so we expect most outliers to lie in this size regime, too. Yet, even though under standard assumptions, solar radiation pressure and rocket forces are not strong enough to produce the measured downward accelerations, small particles may have a much higher water ice content than larger ones, which they would also lose more quickly to sublimation \citep{markkanenThermophysicalModelIcy2020}. The observed trend might thus also be a sign of asymmetric outgassing in the anti-solar direction.

            Finally, we used the extrapolation endpoints shown in Figure~\ref{fig:merged_tracking_results} to roughly estimate the particle ejection velocities. As stated in the caption, these endpoints do not mark the exact ejection times or places, but simply the points where the extrapolated tracks are closest to the ellipse centers. This choice is fairly arbitrary, but since the individual particle origins are impossible to locate precisely, the endpoints nevertheless serve as useful references to roughly estimate when the particles were ejected and at what speed. Accordingly, we find that the initial particle velocities were likely distinctly nonzero. The median initial velocities obtained for the four sequences are also notably similar, centering around $\approx 0.59$\,m\,s$^{-1}$ (STP087: 0.68\,m\,s$^{-1}$, STP088: 0.56\,m\,s$^{-1}$, STP089: 0.45\,m\,s$^{-1}$, STP090: 0.65\,m\,s$^{-1}$; averaged over the respective particle groups). This supports the notion that particles of such sizes are only weakly affected by gas drag, and indicates that they instead gained most of their speed from the initial ejection event. It further agrees well with a growing number of studies that postulate or measured initial particle velocities: \citet{bischoffExperimentsCometaryActivity2019} observed the activity of dust-covered water ice in the lab and find that millimeter-sized particles were ejected with a nonzero initial velocity; \citet{lemosDistributionDynamicsDecimetresized2023, lemosEjectionDynamicsAggregates2024} modeled particle tracks to recreate OSIRIS images of (mostly centimeter-sized) particles recorded farther out in the coma of 67P, and find that the simulated particles required an initial velocity of about 1\,m\,s$^{-1}$ to match the observations; \citet{shiDiurnalEjectionBoulder2024} analyzed the diurnal ejection of boulder clusters from a common source region on 67P's nucleus, and find that their median initial velocity is likely around 0.5\,m\,s$^{-1}$; and \citet{kwonComaEnvironmentComet2023} investigated the dust coma of comet \object{C/2017 K2} and find that their models necessitate nonzero initial particle velocities to reproduce part of the observed coma structure. It therefore seems that the ejection mechanism must be considerably more energetic than a slow or gradual liftoff \citep[see also, e.g.,][]{yelleFormationJetsComet2004, huebnerHeatGasDiffusion2006, beltonCometaryActivityActive2010, kramerPrevailingDustTransportDirections2015, kramerOriginInnerComa2016, knollenbergMiniOutburstNightside2016, vincentAreFracturedCliffs2016, wesolowskiSelectedMechanismsMatter2020}.

        \subsection{Dust coma simulations}
        \label{sect:dust_coma_simulations}
        
            To put our observations into perspective, we used an enhanced version of the water and dust coma modeling software by \citet{marschallDusttoGasRatioSize2020} to simulate the particle dynamics around 67P and its diffuse coma during the time of our observations. The enhanced version of this software now also takes solar radiation pressure, centrifugal force, Coriolis force, and solar tides into account, and allows us to give particles an initial velocity normal to the shape model's surface facet from which they are released. The initial velocities are drawn from a Maxwell-Boltzmann distribution defined via its peak value, $v_\text{init}$ (which is the value we refer to as initial velocity in the following), and are assigned to all particles indiscriminately, even if the modeled gas pressure is theoretically not strong enough to lift them.

            For each of the four sequences, we chose the static gas solution that best matched their respective observational conditions (see Fig.~\ref{fig:gas_models}). The solutions were all computed for epoch 16 of \citet{marschallDusttoGasRatioSize2020}, which covers the time period from Dec. 7, 2015 to Jan. 12, 2016, when 67P was roughly $1.98$\,AU from the Sun at a sub-solar latitude of $-18.2\degree$ (which is right at the center of the sequences' latitude value range, see Table~\ref{tab:meta_data}). The solution setups only differ in sub-solar longitude, where we used $150\degree$ for sequence STP087, $210\degree$ for sequences STP088 and STP089, and $240\degree$ for sequence STP090. All three values almost exactly coincide with the centers of the covered sub-solar latitude value ranges of the respective sequences (see Table~\ref{tab:meta_data}).
     
            Based on these three gas solutions, we then simulated nine different scenarios for each of our sequences. The scenarios are defined by every combination of three different initial velocities (0, 0.25, and 0.5\,m\,s$^{-1}$) and three different activity distributions: only local (i.e., Khonsu, Atum, and Anubis), global, and nonlocal (i.e., global without local activity). For each of these runs, we simulated spherical particles in 17 different size bins distributed logarithmically over a radius range from $10^{-8}$ to 1\,m, where each particle has the same bulk density as the nucleus \citep[533\,kg\,m$^{-3}$,][]{patzoldHomogeneousNucleusComet2016}. To reduce computation time however, we only allowed for one particle to be emitted per model facet and size regime.\footnote{Because of this, especially for flat particle SFDs, the simulations of the diffuse coma appear a bit patchy (Figs.~\ref{fig:STP090_coma_simulations_Kh-Am-Ab_vinit=0.50}--\ref{fig:STP087_vinit_comparison}), since the software has to interpolate the column densities from sparse data. With large amounts of simulated particles, the patchiness disappears. For the same reason, the coma maps of the drag coefficients in Fig.~\ref{fig:gas_models} show some artifacts where not enough particles passed through the affected cells.} 
    
            Figures~\ref{fig:STP087_dust_simulation}--\ref{fig:STP090_dust_simulation} show the results from the particle trajectory simulations for particle sizes similar to those we obtained from the OSIRIS data. For each sequence, initial velocity, and size bin, we randomly selected (up to) 150 simulated particles that were ejected from within the suspected source regions, and projected their trajectories onto the 2D planes of the corresponding OSIRIS/NAC FOVs \citep[via SpiceyPy, a Python wrapper for SPICE,][]{annexSpiceyPyPythonicWrapper2020}. The trajectories start with the beginning of each sequence (the nucleus shapes in their initial configurations are shown for reference), and unless they leave the FOV, they represent the particle motions over a time period of up to two hours. The simulated particles however do not (and are not intended to) retrace our observations. We merely use these simulated particle ensembles to compare them to our observations regarding their general appearance and statistical properties.
            
            According to Figures~\ref{fig:STP087_dust_simulation}--\ref{fig:STP090_dust_simulation}, the initial velocities are essential for reproducing our observed particle tracks (cf. Fig.~\ref{fig:merged_tracking_results}). Of the tested values, $v_\text{init} = 0.5$\,m\,s$^{-1}$ provides the best results for all sequences, which is remarkably similar to the average value we derived from the OSIRIS data (0.59\,m\,s$^{-1}$). In the case of our model, the initial velocities are also not only necessary for ejecting decimeter-sized particles at all, but also for recreating the observed ejection cones. This indicates that besides the particle speed, the shape of the ejection cones may most notably be affected by the local topography. Without the initial velocity, the simulated trajectories of the larger particles appear to \enquote{bend} predominantly in either the clockwise or counterclockwise direction, an effect caused by the viewing geometry and the rotation of the nucleus. Yet when particles are ejected with enough speed from surface areas that face in the opposite direction, their trajectories also appear to bend the other way. In the case of sequence STP090 however, even an initial velocity of 0.5\,m\,s$^{-1}$ is not enough to reproduce the almost symmetric shape of the observed ejection cone (cf. Figs.~\ref{fig:merged_tracking_results} and \ref{fig:STP090_dust_simulation}). 
     
            Despite the initial velocities however, the overall projected (radial) accelerations and velocities of the simulated particles are still lower than what we measured for the real particles (cf. Figs.~\ref{fig:speeds_vs_radii}, \ref{fig:accelerations_vs_radii}, and ~\ref{fig:STP090_dust_simulation_dynamics}). This indicates that even higher initial velocities, locally higher gas production rates (especially of \ce{CO2}, which was not included in the simulations), or lower particle bulk densities (as discussed in Sect.~\ref{sect:particle_dynamics}), may be necessary to reproduce our observations. We also did not model rocket forces, which might have a noticeable effect.
            
            We additionally used the same simulated data from which we recovered the individual trajectories to generate images of the diffuse dust coma as it appeared in the first image of each sequence (cf. Figs.~\ref{fig:sample_images_stp087_stp088} and \ref{fig:sample_images_stp089_stp090}). In this case, particles from every simulated size bin in the range from $10^{-8}$ to 1\,m contributed. Generally, we find that the simulated images fit our observations well (but because we ran 36 different simulations, we do not present all the results here). As an example, Figures~\ref{fig:STP090_coma_simulations_Kh-Am-Ab_vinit=0.50}--\ref{fig:STP090_coma_comparison} show the dust coma simulations of sequence STP090 for all three activity distributions (local, nonlocal, and global), given $v_\text{init} = 0.5$\,m\,s$^{-1}$. The first three of these figures primarily illustrate how the coma simulations are affected by different particle SFDs, which are modeled according to a power law (see, e.g., Eq.~\ref{eq:continuous_powerlaw} and the discussion in Sect.~\ref{sect:particle_size-frequency_distribution}). By visually comparing the diffuse coma structures in these images to those recorded in the OSIRIS images, we find that power-law indices between 3 and 3.5 best reproduce our observations, across all sequences. This agrees well with the value range that we derived for our real, decimeter-sized particles ($3.4 \pm 0.3$--$3.8 \pm 0.4$, see Sect.~\ref{sect:particle_size-frequency_distribution}), suggesting that a single power-law exponent can describe the SFD of both small and large particles.
            
            By comparing the results from the three different activity distributions across all sequences, we also find further evidence that the locations of our suspected source regions are likely correct. For one, images with only local activity show strong dust features above the suspected source regions similar to our observations. Secondly, images with nonlocal activity show that the space above the suspected source regions is only significantly \enquote{contaminated} by dust features from other areas in the cases where the particle SFDs are steep. Because the corresponding particles are much smaller and faster than the particles that we observed however, their contamination is irrelevant. In the relevant particle size range, only relatively weak dust features from other areas appear above the suspected source regions (see Fig.~\ref{fig:STP090_coma_comparison}). Still, in the case of sequence STP090, such contamination might at least in part explain the missing left side of the simulated ejection cone (cf. Figs.~\ref{fig:merged_tracking_results} and \ref{fig:STP090_dust_simulation}), by creating an optical illusion akin to the jet-like features described by \citet{shiComaMorphologyComet2018}. Another reason might be that our gas solutions are static and do not follow the rotation of the nucleus. Depending on the viewing geometry and on how strongly the particles couple with the gas, this could also noticeably affect the projected trajectories.
            
            Like the trajectory simulations, the coma simulations also match our observations best when the initial velocity is highest, as demonstrated by Figures~\ref{fig:STP087_coma_simulations_global_vinit=0.00}--\ref{fig:STP087_vinit_comparison} (for the case of sequence STP087). Clearly, the initial velocity strongly affects the features generated by the largest particles, which seem to make up an essential part of the simulated diffuse coma. The figures also show that different surface regions require different particle SFDs to best reproduce the observed dust features. In the case of sequence STP087 for example, the features near the suspected source region that radiate toward the top right corner are well described by a power-law index $b \approx 3$, while the features on the opposite side that radiate toward the top left corner are better described by a power-law index of at least 3.5.
            
            Finally, Figure~\ref{fig:STP087_vinit_comparison} shows that our model failed to reproduce the strong diffuse coma features seen in the lower right corner of the first image of sequence STP087 (cf. Fig.~\ref{fig:sample_images_stp087_stp088}). These features may have resulted from night-time activity driven by water ice sublimation and sustained by thermal lag, akin to the sunset jets discussed by \citet{shiSunsetJetsObserved2016}, or possibly even from \ce{CO2} ice sublimation \citep{gerigDeviationsFreeradialOutflow2018, gerigDaysidetonightsideDustComa2020, pinzon-rodriguezEffectThermalConductivity2021}.

    \section{Summary and conclusions}
    \label{sect:summary_and_conclusions}

        We analyzed the dynamics and potential origins of 409 decimeter-sized dust particles that were recorded in four OSIRIS/NAC image sequences of 67P's near-nucleus dust coma between December 16, 2015, and January 6, 2016 (post-perihelion).  After tracking thousands of individual dust particles though these sequences, we identified four concentrated groups of recently ejected particles and traced them back to four suspected source regions on the nucleus surface. This allowed us  not only to examine their potential origins, but also to derive their approximate sizes, speeds, and accelerations. Finally, we compared our observations and results to simulations of 67P's dust coma for further evaluation. 
 
        Although we were limited to tracking particles only in the 2D image plane and not in the full 3D environment, our data analysis provides much evidence that the general locations of the suspected source regions are likely correct:
        
        \begin{enumerate}[(i)]
            \item The particle trajectories form distinct ejection cones that taper toward the centers of the suspected source regions.
            \item Even though the suspected source regions were chosen independently from one another, they turned out to be rather well confined and to strongly overlap.
            \item The suspected source regions contain (or are near) areas where global activity models estimate high surface erosion and gas emission rates \citep{combiSurfaceDistributionsProduction2020, lauterIceCompositionClose2022}.
            \item In particular, the suspected source regions derived from image sequences STP087 and STP089 largely coincide with an area in the Khonsu region for which a lot of activity and surface changes have been documented \citep[e.g.,][]{deshapriyaSpectrophotometryKhonsuRegion2016, hasselmannPronouncedMorphologicalChanges2019}.
            \item Unlike other areas, the suspected source regions were continuously well illuminated during the observational periods and the roughly four hours in local time leading up to them.
            \item Trajectory simulations of particles released from the suspected source regions generally reproduce the observed particle tracks well.
            \item Comparisons between different simulations of the diffuse dust coma (local vs. nonlocal) show that most of the simulated dust features above the suspected source regions come from local activity.
        \end{enumerate}
        
        Throughout this paper, we drew several conclusions regarding the nature of the observed activity:

        \begin{enumerate}[(i)]
            \item Instead of homogeneous activity, the ejection of large particles ($\gtrapprox 1$\,cm) can occur distinctly localized.
            \item The concentrated ejection of large particles does not necessarily correlate (in strength, location, or orientation) with that of small particles ($\lessapprox 1$\,cm) that make up the diffuse coma. This may be evidence that water-driven erosion and \ce{CO2}-driven ejection of large chunks cannot happen simultaneously at the same location.
            \item The suspected source regions of the particles that we traced back to the nucleus surface predominantly lie in the Khonsu-Atum-Anubis area, and the observed activity may be linked to rugged terrain or steep slopes like scarps, cliffs, or fractures.
            \item The studied particles range in size from about 5\,cm to 1.15\,m in (equivalent) radius. Power-law fits to their SFDs best describe the data with power-law indices between $3.4 \pm 0.3$ and $3.8 \pm 0.4$. This indicates that shortly after ejection, most of the mass is still contained in the larger particles, although ultimately most of them likely did not escape the nucleus gravity. The index values also agree notably well with those obtained for submillimeter-sized particles \citep[3.7 and $3.1\pm0.5$,][]{fulleEvolutionDustSize2016, merouaneDustParticleFlux2016}, and might reflect an SFD transition of the surface material located in the suspected source regions \citep{deshapriyaSpectrophotometryKhonsuRegion2016, hasselmannPronouncedMorphologicalChanges2019}. 
            \item Solar irradiation alone cannot explain the locality of the observed activity. Additionally, surface accelerations in the suspected source regions are relatively high, ruling out gravity and centrifugal forces as decisive factors as well.
            \item The projected radial particle velocities directly provide upper limit estimates for the particle fall-back fractions, which lie between $75$ and $91$\%. The data indicate that essentially all particles larger than 40\,cm likely fell back onto the nucleus surface. 
            \item The distributions of the projected radial particle accelerations as functions of the particle radii are well described by the local surface acceleration (lower bound) and gas drag (upper bound). The gas drag parameters, however, are degenerate and cannot be precisely constrained. Values from our water and dust coma simulations nevertheless indicate that the local gas production rate was likely several times higher ($Q_\text{g} = 3.6 \cdot 10^{-5}$\,kg\,s$^{-1}$\,m$^{-2}$) than the prediction by our purely insolation-driven model and the peak average production rates estimated by \citet{lauterIceCompositionClose2022}. 
            \item Some particles exhibit downward accelerations that are much stronger than the local surface accelerations. Most of these outliers are likely caused by inaccurate measurements and statistical effects, but their general trend might also be a sign of asymmetric outgassing. 
            \item Rough estimates of the initial particle velocities are distinctly nonzero and average around 0.59\,m\,s$^{-1}$, which indicates that the particles likely gained most of their speed from the initial ejection event.
            \item Our dynamics simulations of decimeter-sized particles in the coma of 67P support the need for higher local activity to reproduce the observed trajectories. Simulated particles larger than $\approx 32$\,cm could not be lifted from the suspected source regions without introducing initial velocities in addition to gas drag. Even with an initial velocity of $\approx 0.5$\,m\,s$^{-1}$ the simulated particles were generally still slower than those we observed. 
            \item The inclusion of initial velocities also shows that they are necessary for reproducing the observed ejection cones, which indicates that the local topography plays an important role in shaping these cones.
            \item Both, the simulated dynamics of individual particles, and the simulated images of the diffuse dust coma, match the observations best when the initial velocity is the highest ($\approx 0.5$\,m\,s$^{-1}$). This is further evidence that initial velocities are an essential aspect of the ejection process.
            \item The simulated images additionally reproduce our observations best given particle SFDs described by power laws with indices equal to 3 or 3.5. This agrees well with the value range that we obtained from our real particle populations, but we also found that some dust features require different size distributions to be well reproduced.
        \end{enumerate}
        
        Overall, our observational and modeling results strongly suggest that the concentrated local ejection of decimeter-sized particles cannot be explained with water-driven activity and favorable illumination conditions alone. Instead, the composition and structure of the suspected source regions seem to be the deciding factors; of these, we deem an overabundance of volatiles, in particular of \ce{CO2} ice, to be the most probable cause. This is in line with current particle ejection models \citep[e.g.,][]{gundlachActivityCometsUnderstanding2020, fulleHowCometsWork2020, wesolowskiSelectedMechanismsMatter2020, ciarnielloMacroMicroStructures2022, davidssonCO2drivenSurfaceChanges2022}, which necessitate the sublimation of \ce{CO2} ice in deeper surface layers to eject decimeter-sized particles. Additionally, our results show that decimeter-sized particles are very likely ejected with substantial nonzero initial velocities, which agrees well with other recent studies \citep[e.g.,][]{bischoffExperimentsCometaryActivity2019, lemosDistributionDynamicsDecimetresized2023, lemosEjectionDynamicsAggregates2024, kwonComaEnvironmentComet2023, shiDiurnalEjectionBoulder2024}, and implies that the ejection mechanism must be considerably more energetic than a slow or gradual liftoff.

    \begin{acknowledgements}
        We thank the anonymous referee for their valuable feedback; Xian Shi, Nicholas Attree, Yuri Skorov, Marco Fulle, and Asmus Freitag for (proof)reading our manuscript and providing helpful comments; Miryam Merk, for statistical consulting; Sonia Fornasier, Matthias Läuter, and Tobias Kramer for providing their data; Aaron Clauset for discussing the intricacies of power-law fitting with us; and Pedro Hasselmann, Maurizio Pajola, Mohamed Ramy El-Maarry, Nicolas Thomas, Frank Preusker, Michael Combi, Koji Wada, Jean-Baptiste Vincent, Johannes Markkanen, and Giovanna Rinaldi for providing tools and discussing certain aspects of our paper.
        \newline
        We acknowledge the operation and calibration team at MPS and the Principal Investigator Holger Sierks on behalf of the OSIRIS Team for providing the OSIRIS images and related data sets. OSIRIS was built by a consortium of the Max-Planck-Institut für Sonnensystemforschung, Göttingen, Germany; the CISAS University of Padova, Italy; the Laboratoire d'Astrophysique de Marseille, France; the Instituto de Astrofísica de Andalucia, CSIC, Granada, Spain; the Research and Scientific Support Department of the European Space Agency, Noordwijk, The Netherlands; the Instituto Nacional de Técnica Aeroespacial, Madrid, Spain; the Universidad Politéchnica de Madrid, Spain; the Department of Physics and Astronomy of Uppsala University, Sweden; and the Institut für Datentechnik und Kommunikationsnetze der Technischen Universität Braunschweig, Germany. The support of the national funding agencies of Germany (DLR), France (CNES), Italy (ASI), Spain (MEC), Sweden (SNSB), and the ESA Technical Directorate is gratefully acknowledged. We thank the Rosetta Science Ground Segment at ESAC, the Rosetta Missions Operations Centre at ESOC and the Rosetta Project at ESTEC for their outstanding work enabling the science return of the Rosetta Mission.
        \newline
        MP, JA, and PL acknowledge funding by the ERC Starting Grant No. 757390 Comet and Asteroid Re-Shaping through Activity (CAstRA). JA acknowledges funding by the Volkswagen Foundation. MP and PL conducted the work in this paper in the framework of the International Max-Planck Research School (IMPRS) for Solar System Science at the University of Göttingen.

    \end{acknowledgements}

    \bibliographystyle{aa}

    \begin{appendix}

        \twocolumn[\section{Sample images from sequences STP089 and STP090}
        \label{sect:sample_images}
        \vspace*{3pt}
        
            {\centering\includegraphics[width=\textwidth]{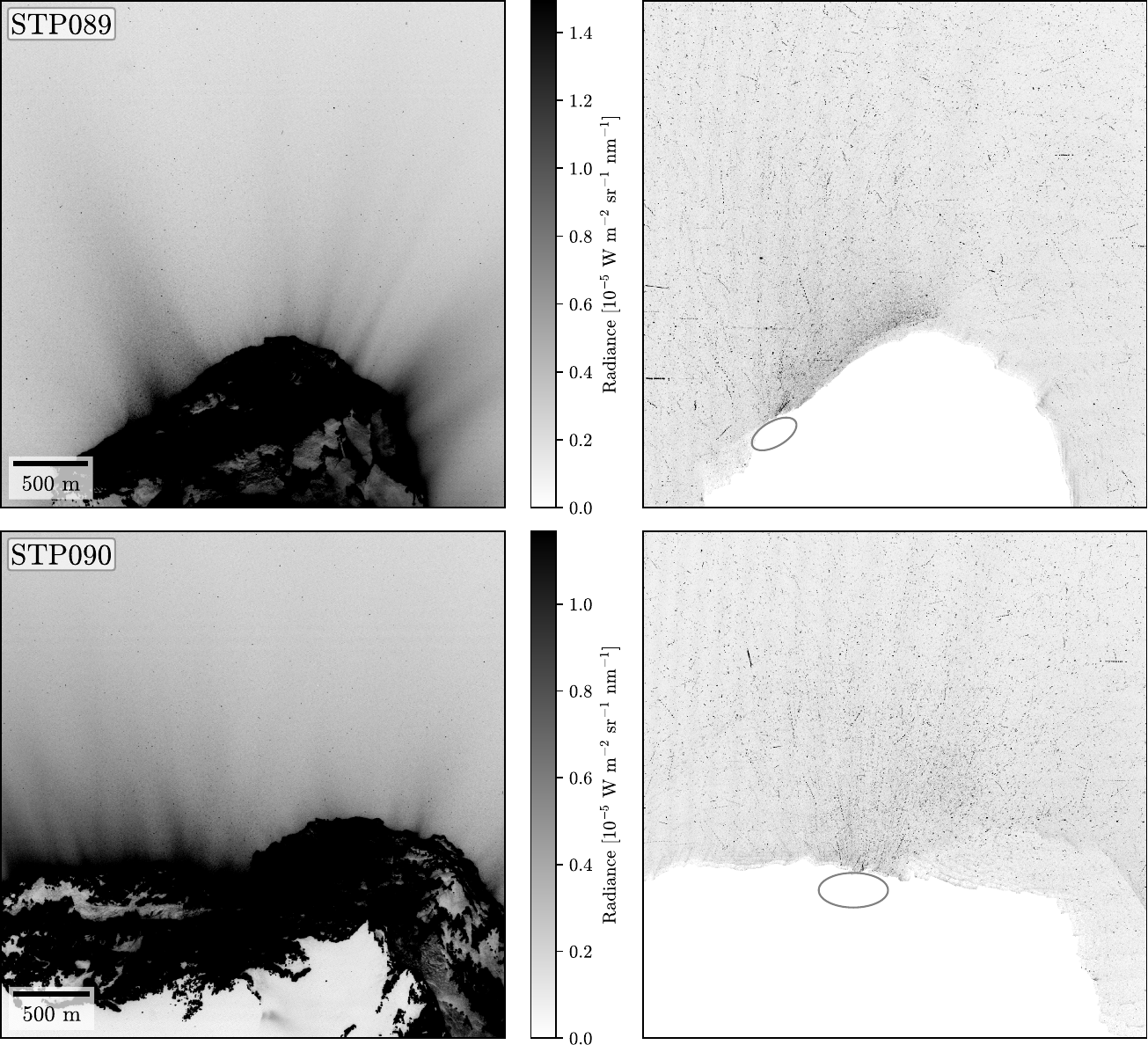}}
            \begin{minipage}{\textwidth}
                \captionof{figure}{Sample images from sequences STP089 and STP090. The first images of the respective sequences are on the left, the master images on the right. The ellipses in the master images mark the suspected source regions of the concentrated particle groups. All images are brightness-inverted and had their contrasts improved for better readability (continuation of Fig.~\ref{fig:sample_images_stp087_stp088}).}
                \label{fig:sample_images_stp089_stp090}
                \vspace*{12pt}
            \end{minipage}]
            \FloatBarrier

        \section{Reasons for rejection of image sequences}
        \label{sect:rejection_reasons}
        
            In the following, we list the reasons why we excluded other OSIRIS image sequences from our analysis (roughly from most to least critical): 
            \begin{itemize}
                \item An incomplete pair of subsequences (i.e., only one exists) or an insufficient number of images (many subsequences consist of less than twenty images). If the covered time period is long, subsequent particle detections can lie far apart, which makes particle tracking very difficult, especially if there is no \enquote{stem} of detections linked over a short interval (cf. Fig.~\ref{fig:STP089_timeline}), that provides accurate predictions. Conversely, if the covered time period is short, the derived particle dynamics can be unreliable because the tracks do not evolve enough for fits to be resistant against smaller deviations like pointing fluctuations or the inclusion of unrelated detections.
                \item Binning, which severely hampers particle detection.
                \item A lack of sidereal objects, which are needed to correct for pointing fluctuations.
                \item An abrupt and substantial change in (commanded) camera pointing, which is nontrivial to correct for, and makes visual confirmation of particle tracks mostly impossible.
                \item Long time gaps/periods, which make the continuous tracking of the same particles difficult and result in many particles to have left the FOV.
                \item No concentrated group(s) of particles that seemingly originate from the same surface area (see Sect.~\ref{sect:track_seletion} as to why this is important).  
                \item A low number of (reliable) tracks (e.g., due to a low number of particles). 
                \item Nucleus outside the FOV, which makes associating particles with potential source regions on its surface much more speculative.
                \item Too large nucleocentric distances of the spacecraft, which do not allow for particles near the nucleus to be distinguishable from the diffuse coma or associated with potential source regions.
                \item Different time signatures (i.e., images come in singles or triplets instead of pairs, which require at minimum an adaptation of the tracking algorithm).
                \item Missing or an uneven number of images (e.g., in the used calibration levels). 
                \item Defect/artifact-riddled/incomplete images (see, e.g., the artifacts around the nucleus in the first image of sequence STP088, shown in Fig.~\ref{fig:sample_images_stp087_stp088}).
            \end{itemize}

        \section{Caveats of polynomial fitting}
        \label{sect:fitting_caveats}

            \begin{figure*}[b]
                \centering
                \includegraphics[width=\linewidth]{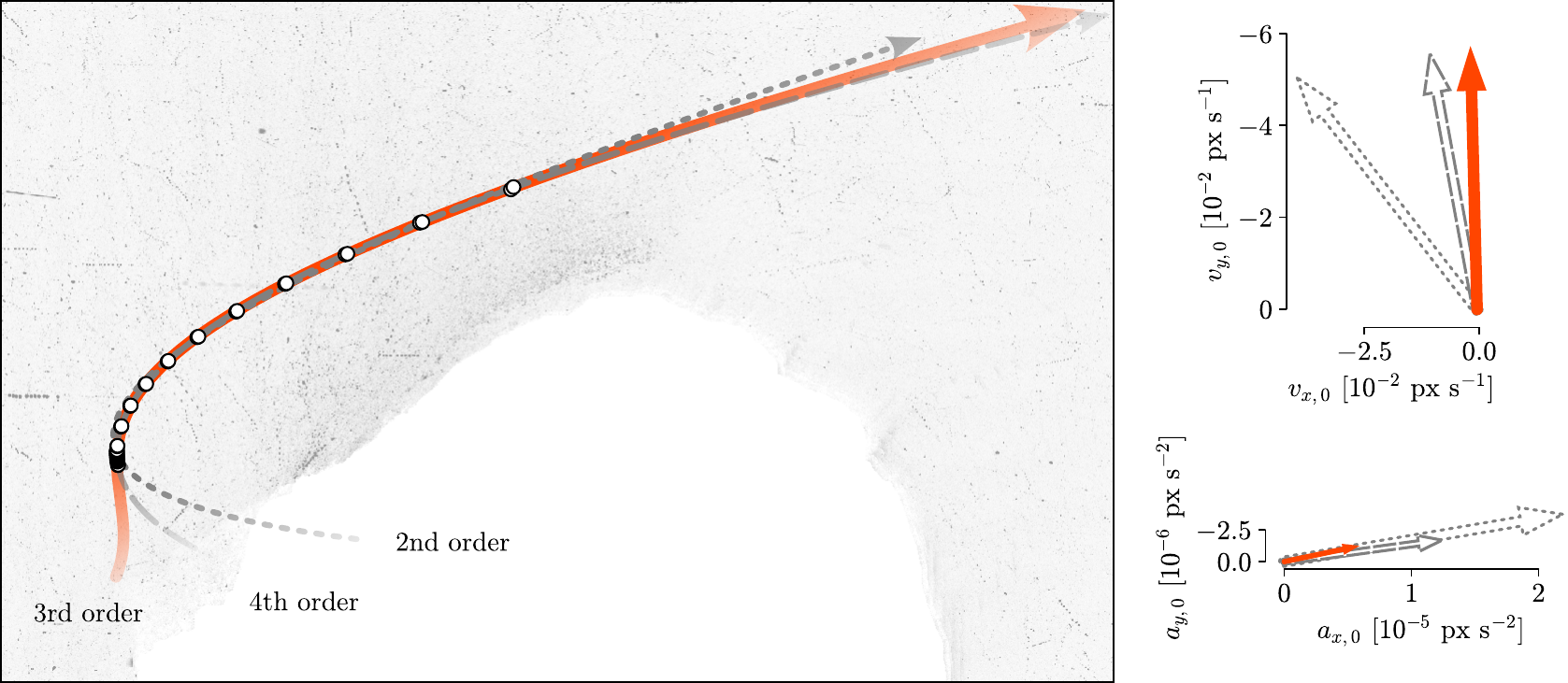}
                \caption{Same particle track shown in Fig.~\ref{fig:sample_track} (white circles), but this time fitted with a second-, third-, and fourth-order polynomial, each extrapolated an hour back and over two and a half hours forward in time. The plots on the right show the respective derived vertical and horizontal particle velocities and accelerations at $t = 0$\,s.}
                \label{fig:fit_order_demonstration}
            \end{figure*}
    
            As Figure~\ref{fig:sample_track} shows, fitting third-order polynomials is sometimes not only required for tracking particles successfully, but often simply the more appropriate choice. From a physical standpoint, third-order polynomials are justified, since the particle acceleration changes over time due to the complex gas and micro-gravity environment and the possibility for asymmetric outgassing. Even fourth-order polynomials may be fair. Yet because of residual pointing fluctuation and other effects, the positional data of our particles are not precise enough to allow for the detection of such delicate signals. In some cases, the order of the fitted polynomial can also substantially affect the derived velocity and acceleration vectors and change the extrapolated course of the track (see Fig.~\ref{fig:fit_order_demonstration} for an extreme example). Our ability to extrapolate particle tracks is thus limited, which is one reason why we only trace back particles for at most half an hour.
            
            To ensure that our statistical results are nonetheless reliable, we tested how much they are affected by the order of the fitted polynomials (second or third). The most notable difference was in the track populations that intersect with the suspected source regions. Some tracks only do so when using a second- but not a third-order polynomial, and others vice versa. The corresponding radius, velocity, and acceleration distributions, however, are very similar, and do not significantly change the derived qualities. The fitted SFD power-law indices, for example, differ by no more than 0.1. Based on this analysis, we consider our statistical results reliable, and our conclusions remain unaffected.

        \onecolumn
        \section{Dust coma simulation results}
        \label{sect:simulation_results}
        
            \begin{figure*}[!ht]
                \centering
                \includegraphics[width=\linewidth]{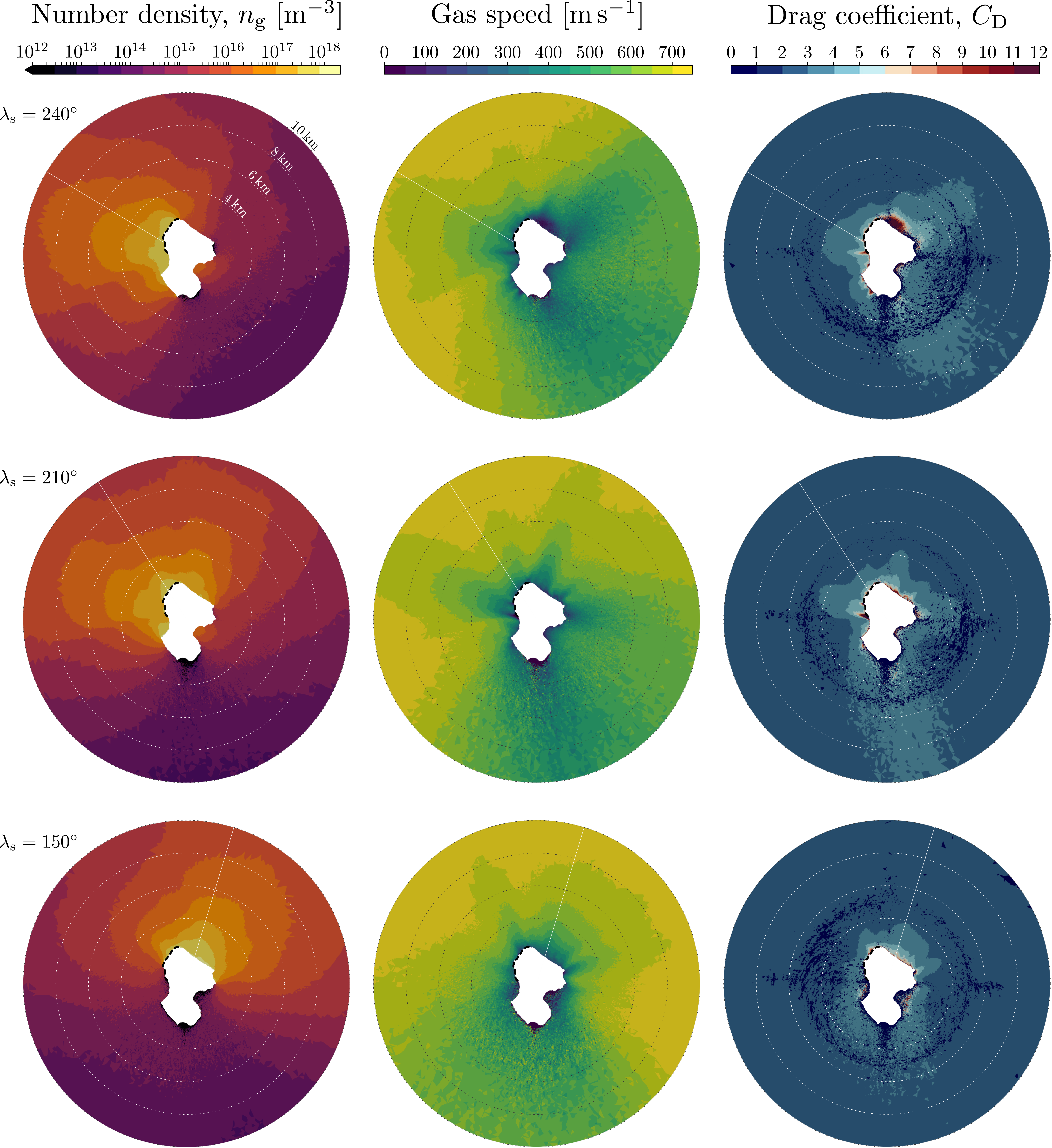}
                \caption{Properties of the three static gas solutions that we used for our dust coma models. The plots show model cross sections that slice through the suspected source regions, which are indicated by the bold dashed curves on the top left side of the nucleus. The solar directions are highlighted by the white lines. The drag coefficient plots show the results computed for global activity of 32 cm  particles, which are representative for the whole relevant size range from 1\,cm to 1\,m. The circular artifacts in these plots, around 3\,km from the nucleus, are a consequence of how the simulation domain was built (with a transition region between an inner sphere with very small cells and and an outer region with much larger cells) and the low number of simulated particles.}
                \label{fig:gas_models}
            \end{figure*}
    
            \begin{figure*}[!htp]
                \centering
                \includegraphics[width=0.975\linewidth]{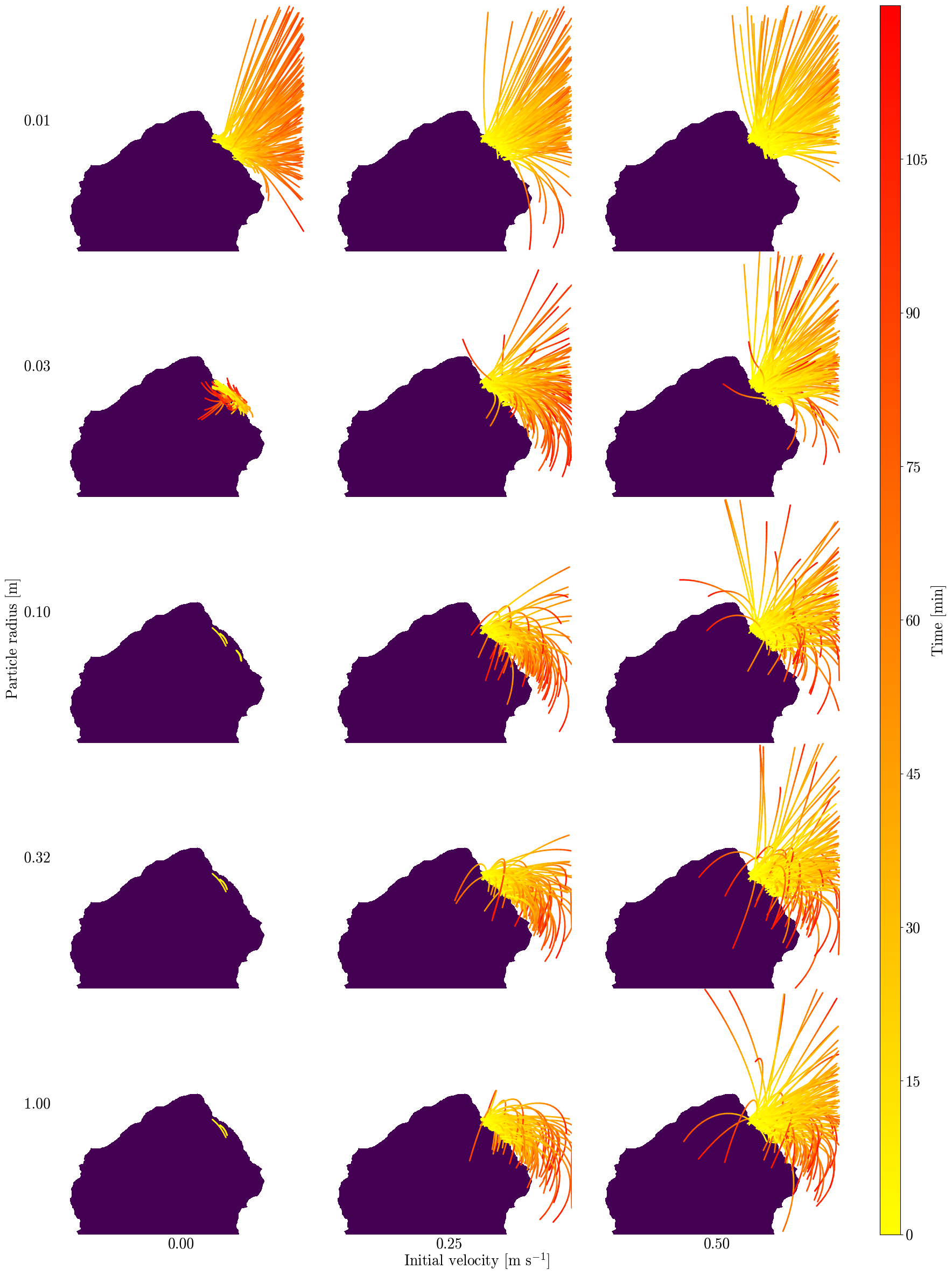}
                \caption{Trajectory simulations of up to 150 randomly selected particles ejected from within the suspected source region of sequence STP087.}
                \label{fig:STP087_dust_simulation}
            \end{figure*} 
            
            \begin{figure*}[!htp]
                \centering
                \includegraphics[width=0.975\linewidth]{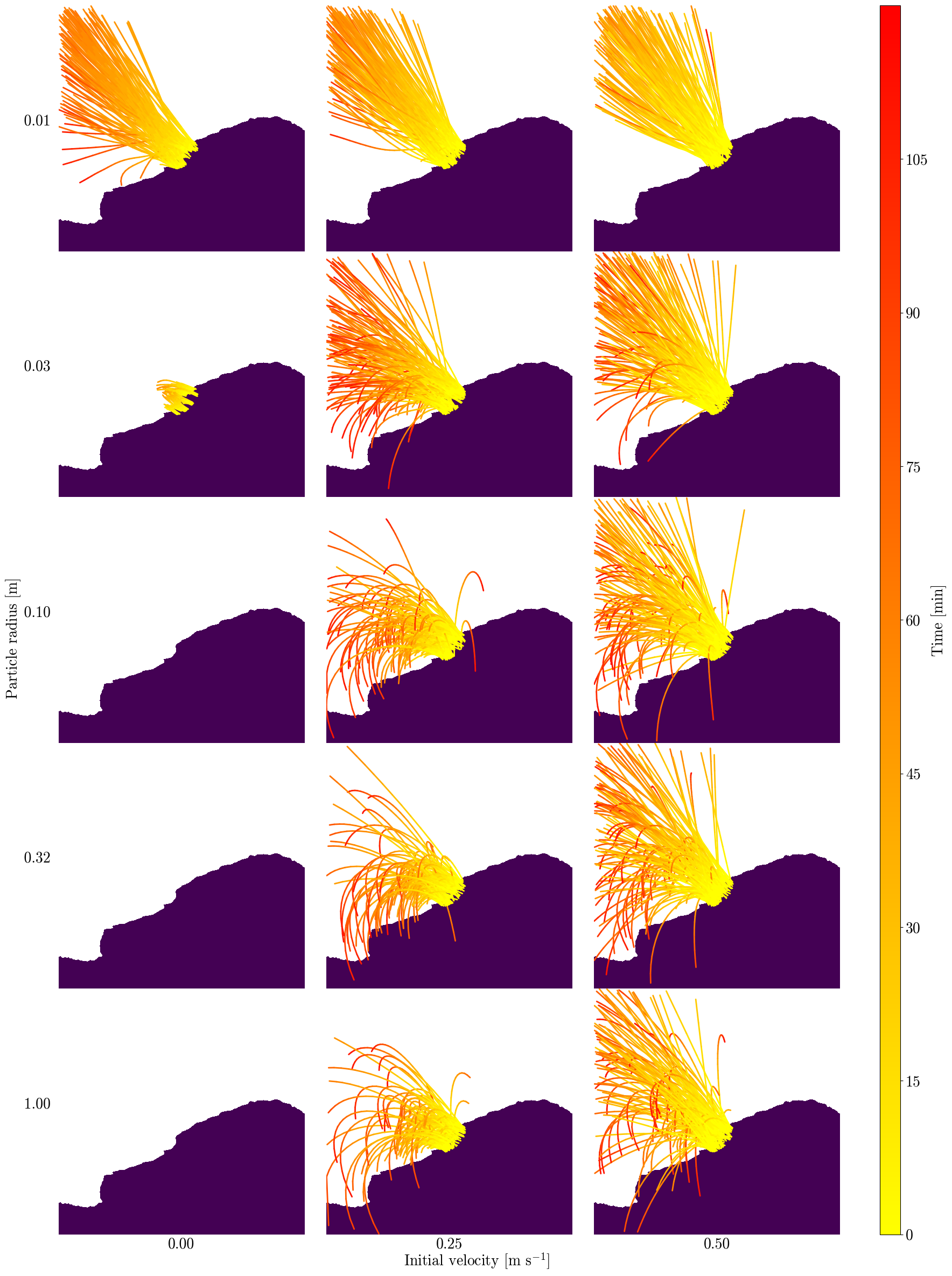}
                \caption{Trajectory simulations of up to 150 randomly selected particles ejected from within the suspected source region of sequence STP088.}
                \label{fig:STP088_dust_simulation}
            \end{figure*} 
            
            \begin{figure*}[!htp]
                \centering
                \includegraphics[width=0.975\linewidth]{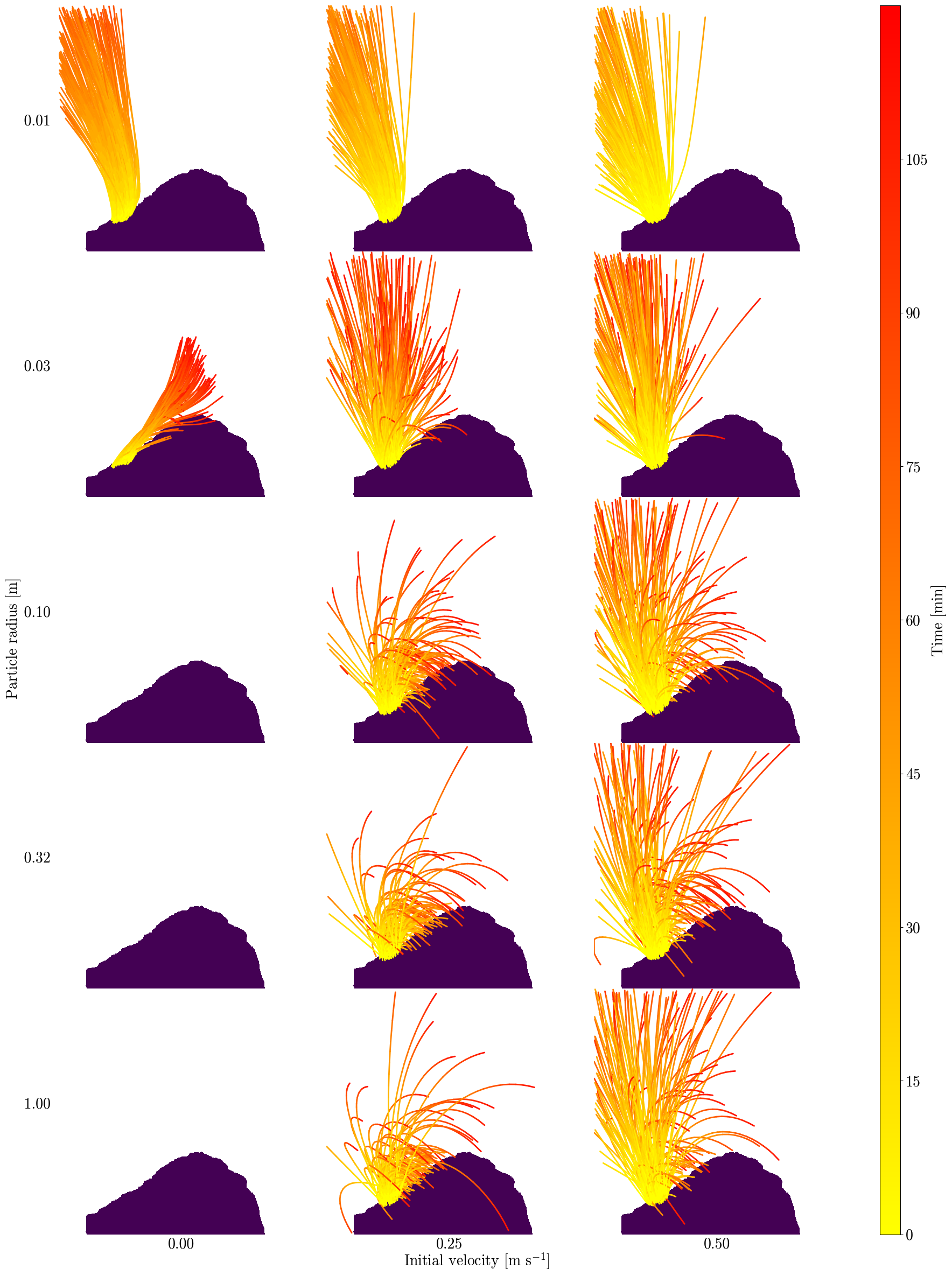}
                \caption{Trajectory simulations of up to 150 randomly selected particles ejected from within the suspected source region of sequence STP089.}
                \label{fig:STP089_dust_simulation}
            \end{figure*} 
            
            \begin{figure*}[!htp]
                \centering
                \includegraphics[width=0.975\linewidth]{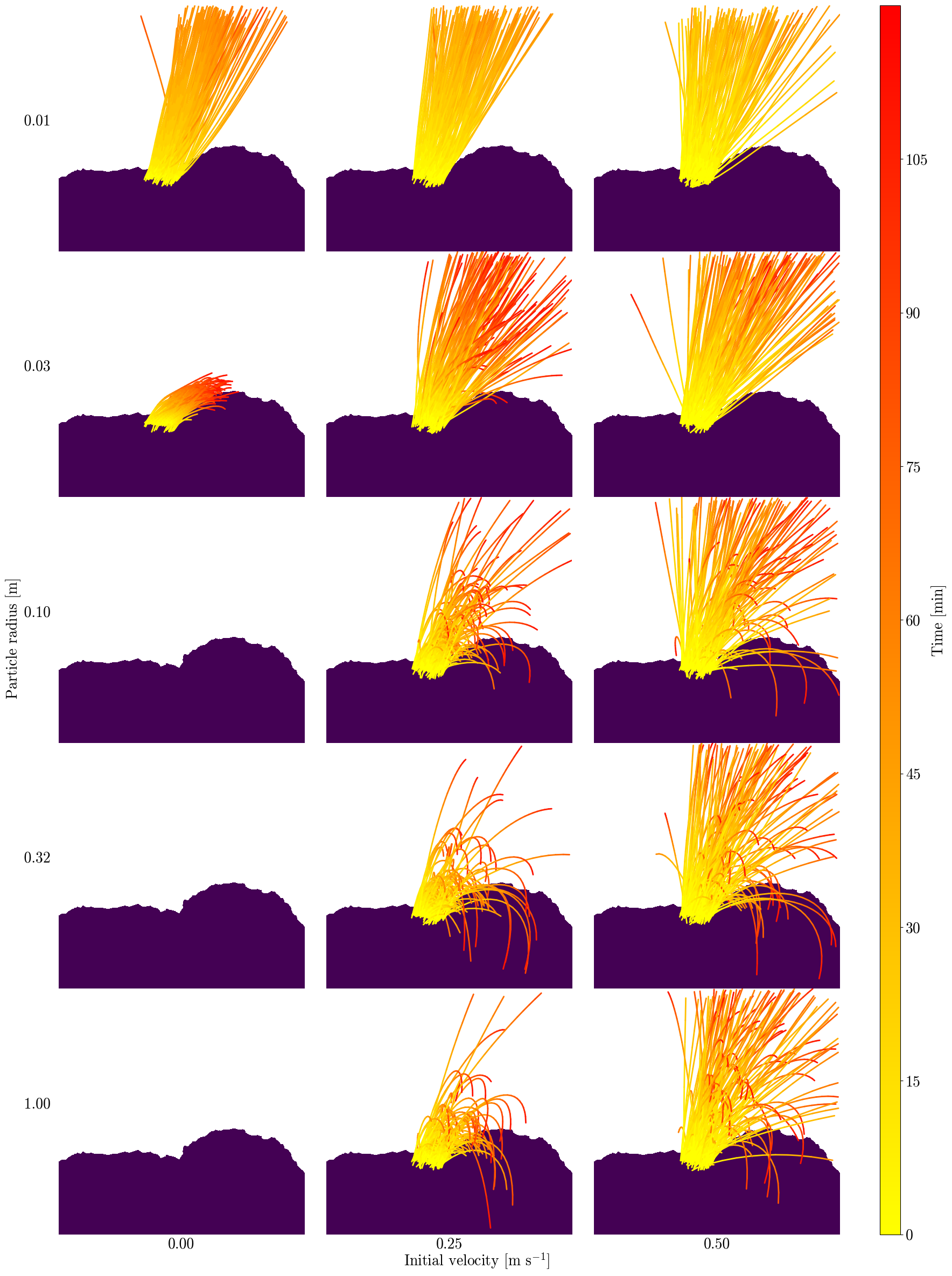}
                \caption{Trajectory simulations of up to 150 randomly selected particles ejected from within the suspected source region of sequence STP090.}
                \label{fig:STP090_dust_simulation}
            \end{figure*} 
            
            \begin{figure*}[!htp]
                \centering
                \includegraphics[width=0.92\linewidth]{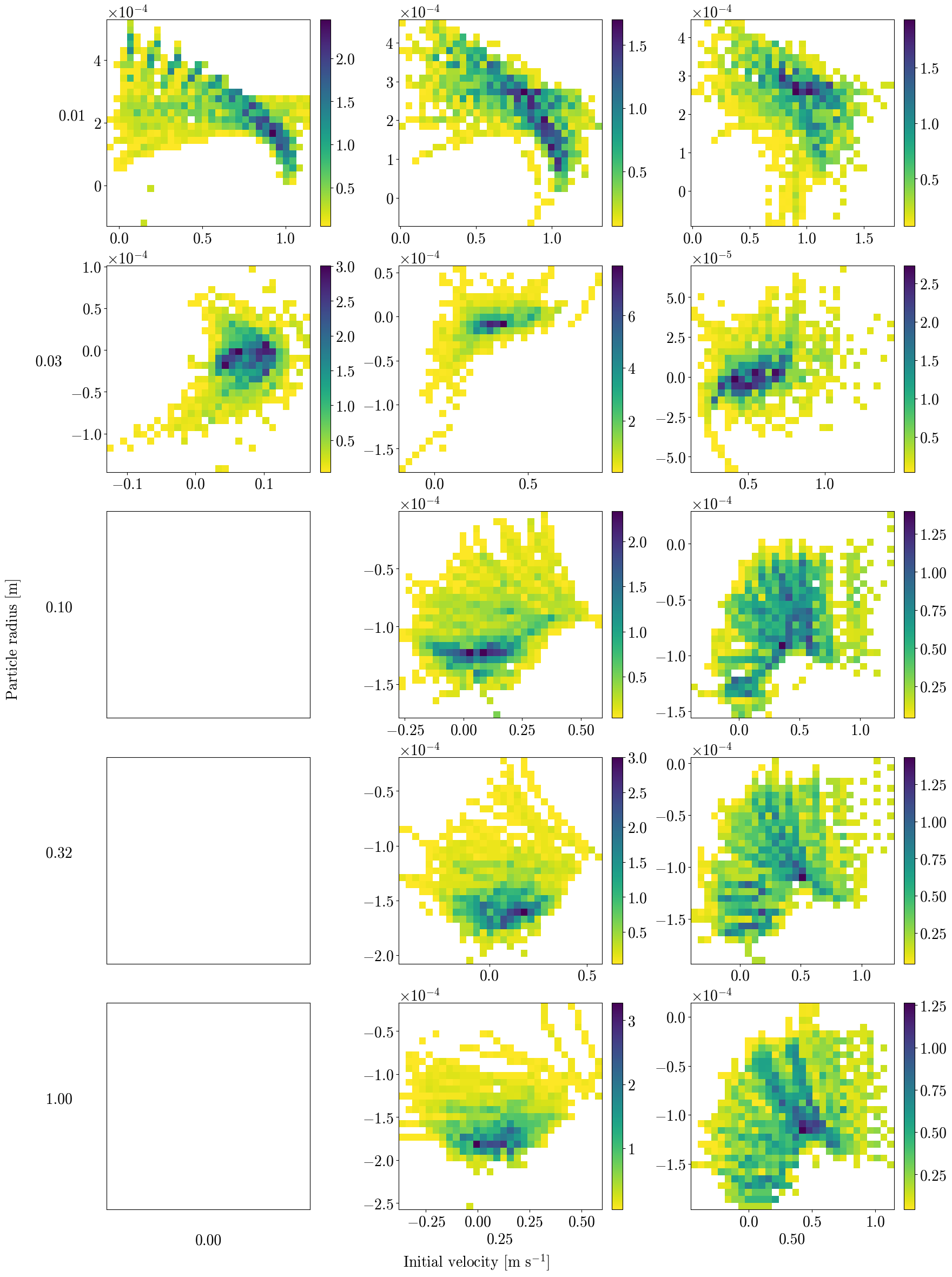}
                \caption{Projected radial accelerations and velocities of the particles shown in Fig.~\ref{fig:STP090_dust_simulation} (sequence STP090). As with the OSIRIS data, the values were determined by fitting third-order polynomials to the projected tracks. In each subplot, the x-axis shows the projected radial velocity in m\,s$^{-1}$, the y-axis the projected radial acceleration in m\,s$^{-2}$, and the color bar the number of measurements, which are again weighted by the number of measurements per respective track (residence time weighting). The three bottom left panels are blank because no particles were ejected in these instances.}
                \label{fig:STP090_dust_simulation_dynamics}
            \end{figure*}

            \begin{figure*}[!ht]
                \centering
                \includegraphics[width=0.62\linewidth]{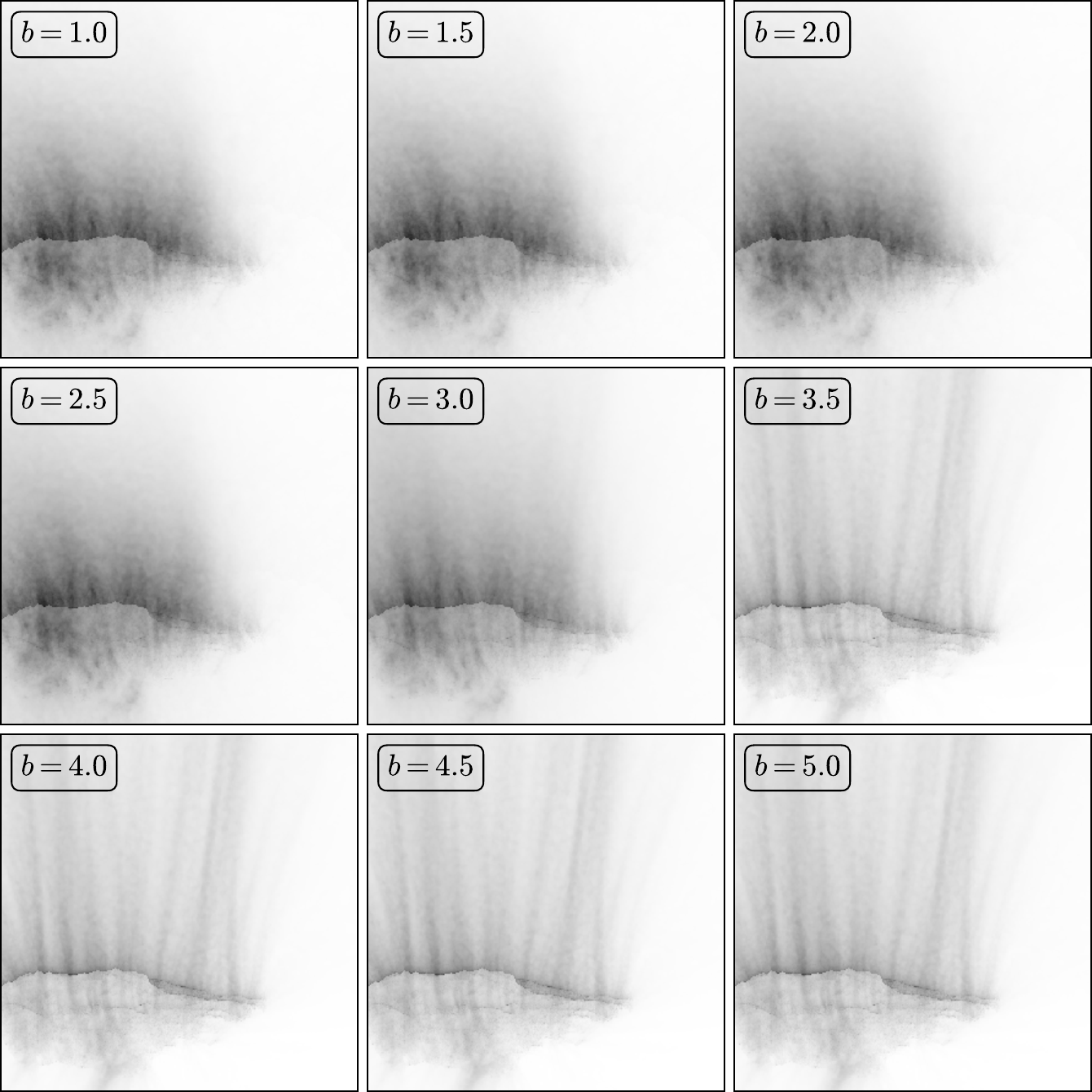}
                \caption{Coma simulations for sequence STP090 as a function of the particle SFD power-law index, $b$, given local activity and $v_\text{init} = 0.5$\,m\,s$^{-1}$.}
                \label{fig:STP090_coma_simulations_Kh-Am-Ab_vinit=0.50}
            \end{figure*}
            
            \begin{figure*}[!ht]
                \centering
                \includegraphics[width=0.62\linewidth]{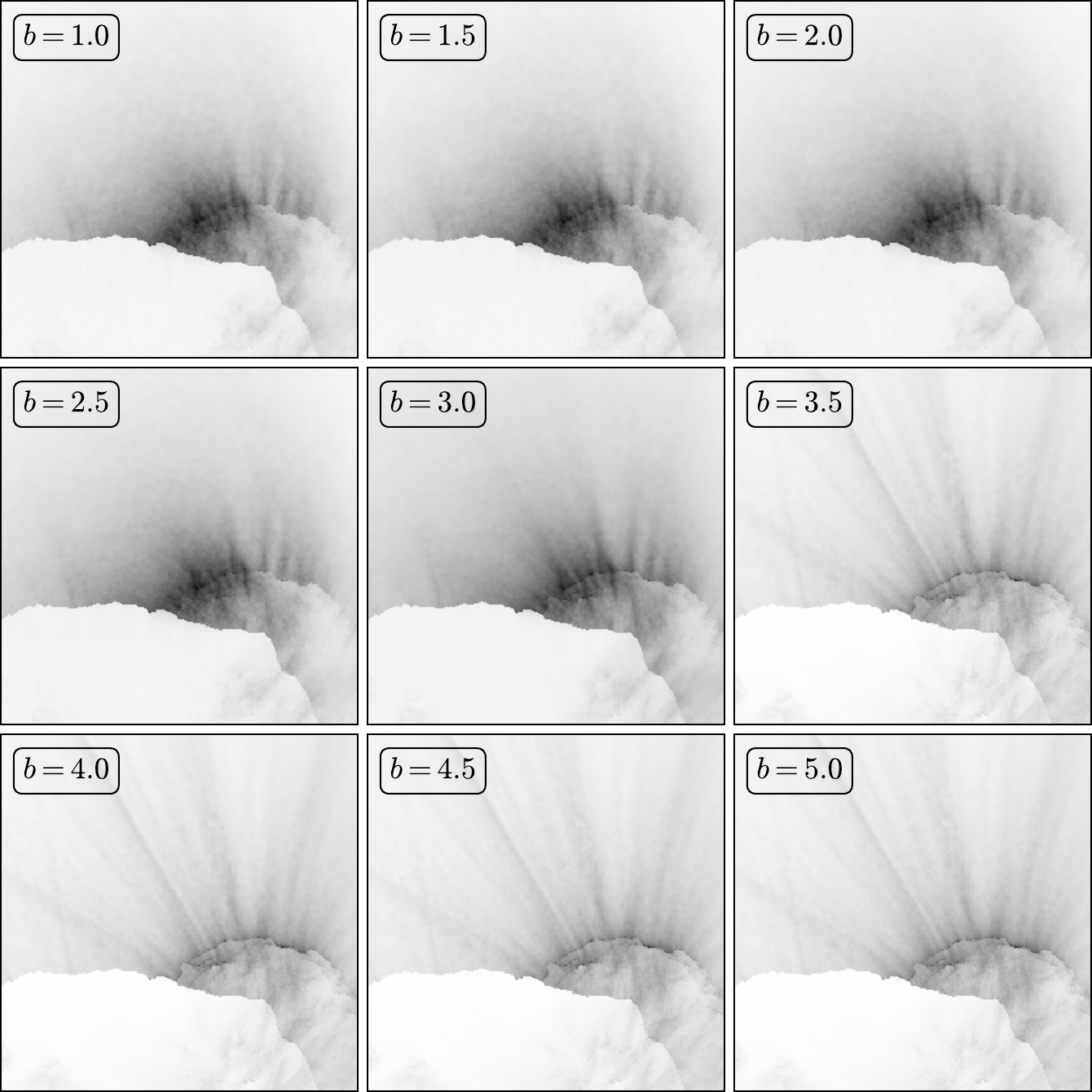}
                \caption{Coma simulations for sequence STP090 as a function of the particle SFD power-law index, $b$, given nonlocal activity and $v_\text{init} = 0.5$\,m\,s$^{-1}$.}
                \label{fig:STP090_coma_simulations_all-except_vinit=0.50}
            \end{figure*}
            
            \begin{figure*}[!ht]
                \centering
                \includegraphics[width=0.59\linewidth]{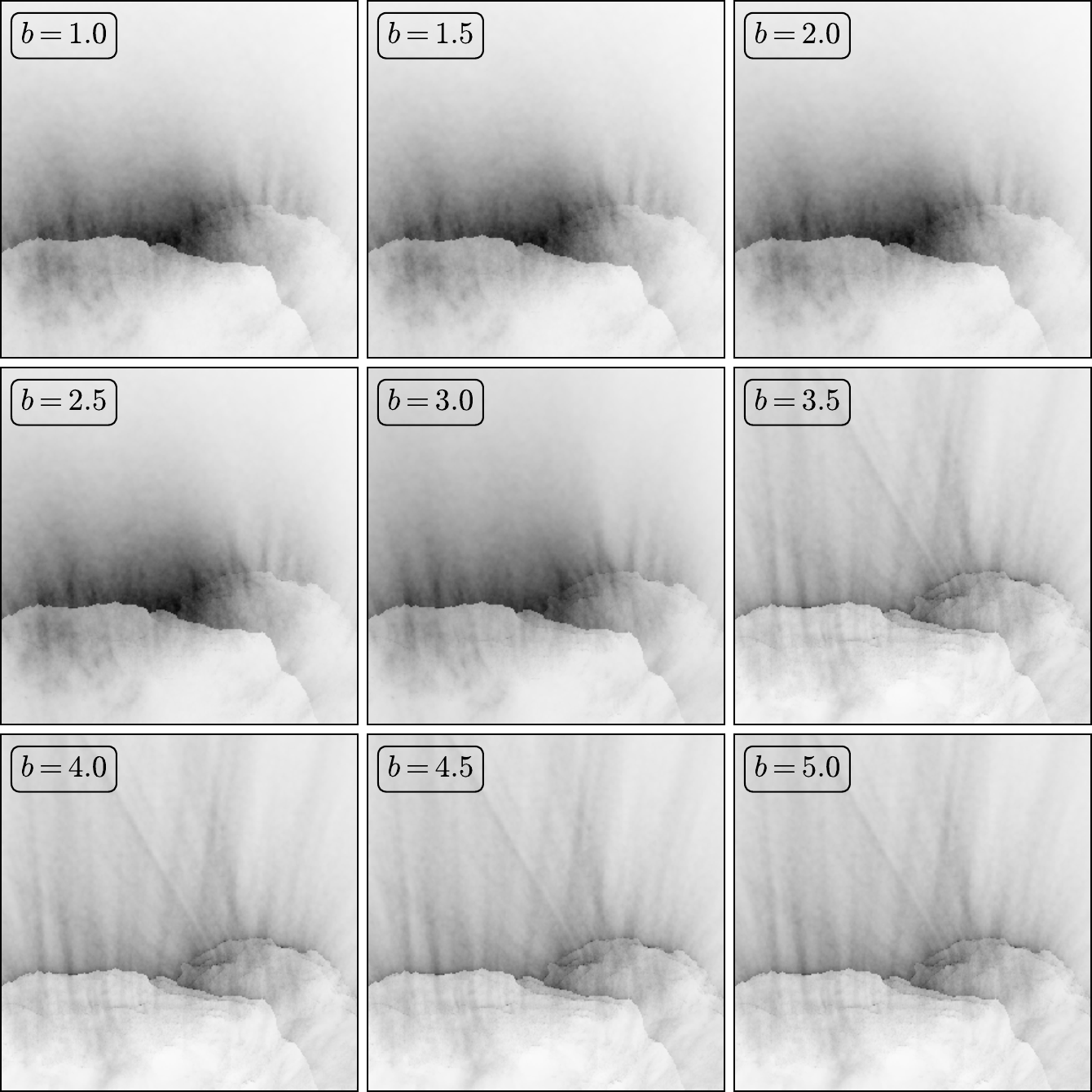}
                \caption{Coma simulations for sequence STP090 as a function of the particle SFD power-law index, $b$, given global activity and $v_\text{init} = 0.5$\,m\,s$^{-1}$.}
                \label{fig:STP090_coma_simulations_global_vinit=0.50}
            \end{figure*}
            
            \begin{figure*}[!ht]
                \centering
                \includegraphics[width=0.59\linewidth]{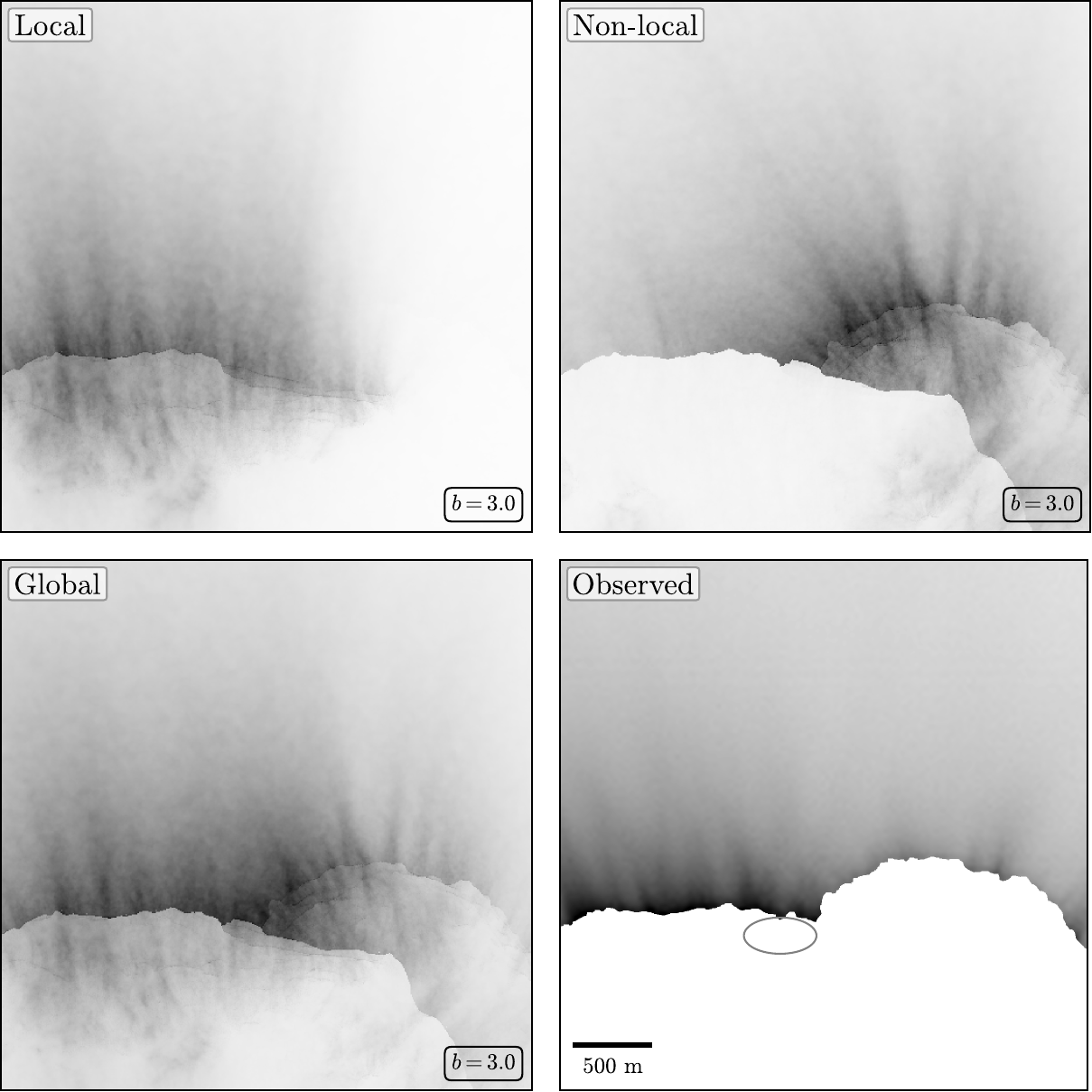}
                \caption{Comparison between coma simulations with different activity distributions given $v_\text{init} = 0.5$\,m\,s$^{-1}$, and the observed coma in the first image of sequence STP090. The image of the observed coma is the background signal that we subtracted during the preparation for the tracking procedure (see Sect.~\ref{sect:data_selection} and \citealp{pfeiferTrailCometTail2022}). The ellipse indicates the suspected source region. All images are brightness-inverted and had their contrasts improved individually for better reading. Because of this, the absolute intensity levels should not be compared across images, but only relative to other areas of the same image.}
                \label{fig:STP090_coma_comparison}
            \end{figure*}

            \begin{figure*}[!ht]
                \centering
                \includegraphics[width=0.62\linewidth]{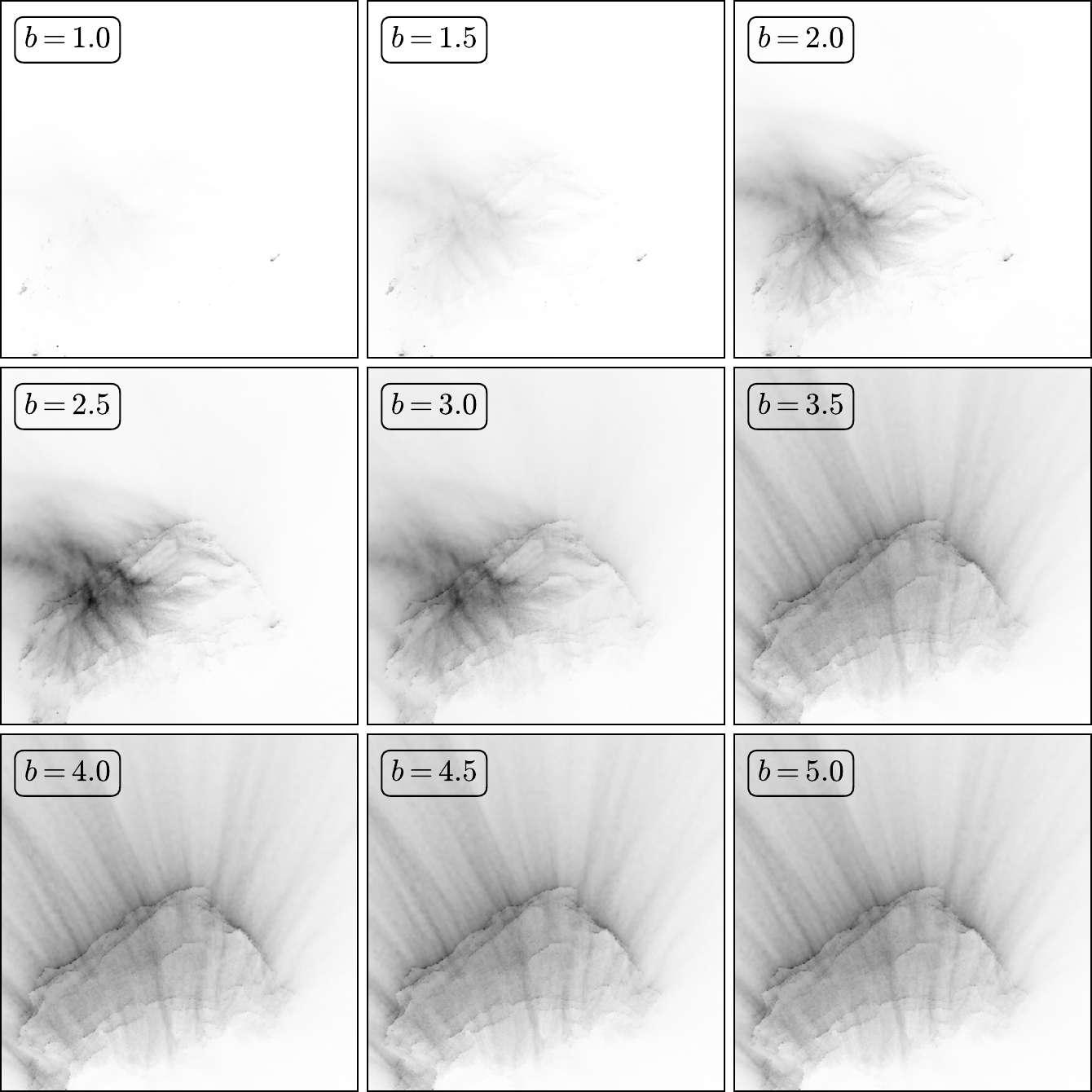}
                \caption{Global coma simulations for sequence STP087 as a function of the particle SFD power-law index, $b$, given $v_\text{init} = 0.0$\,m\,s$^{-1}$.}
                \label{fig:STP087_coma_simulations_global_vinit=0.00}
            \end{figure*}
            
            \begin{figure*}[!ht]
                \centering
                \includegraphics[width=0.62\linewidth]{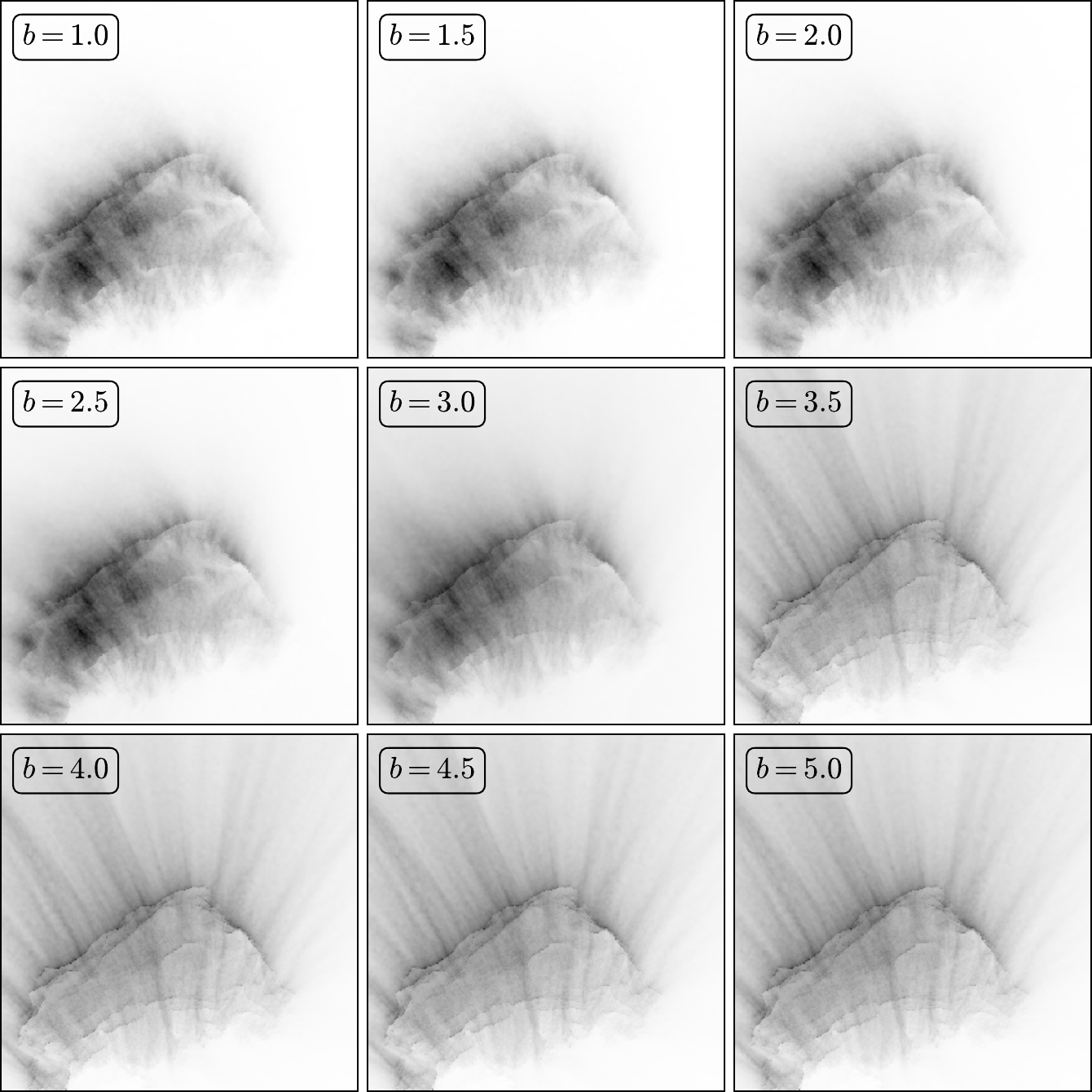}
                \caption{Global coma simulations for sequence STP087 as a function of the particle SFD power-law index, $b$, given $v_\text{init} = 0.25$\,m\,s$^{-1}$.}
                \label{fig:STP087_coma_simulations_global_vinit=0.25}
            \end{figure*}
            
            \begin{figure*}[!ht]
                \centering
                \includegraphics[width=0.59\linewidth]{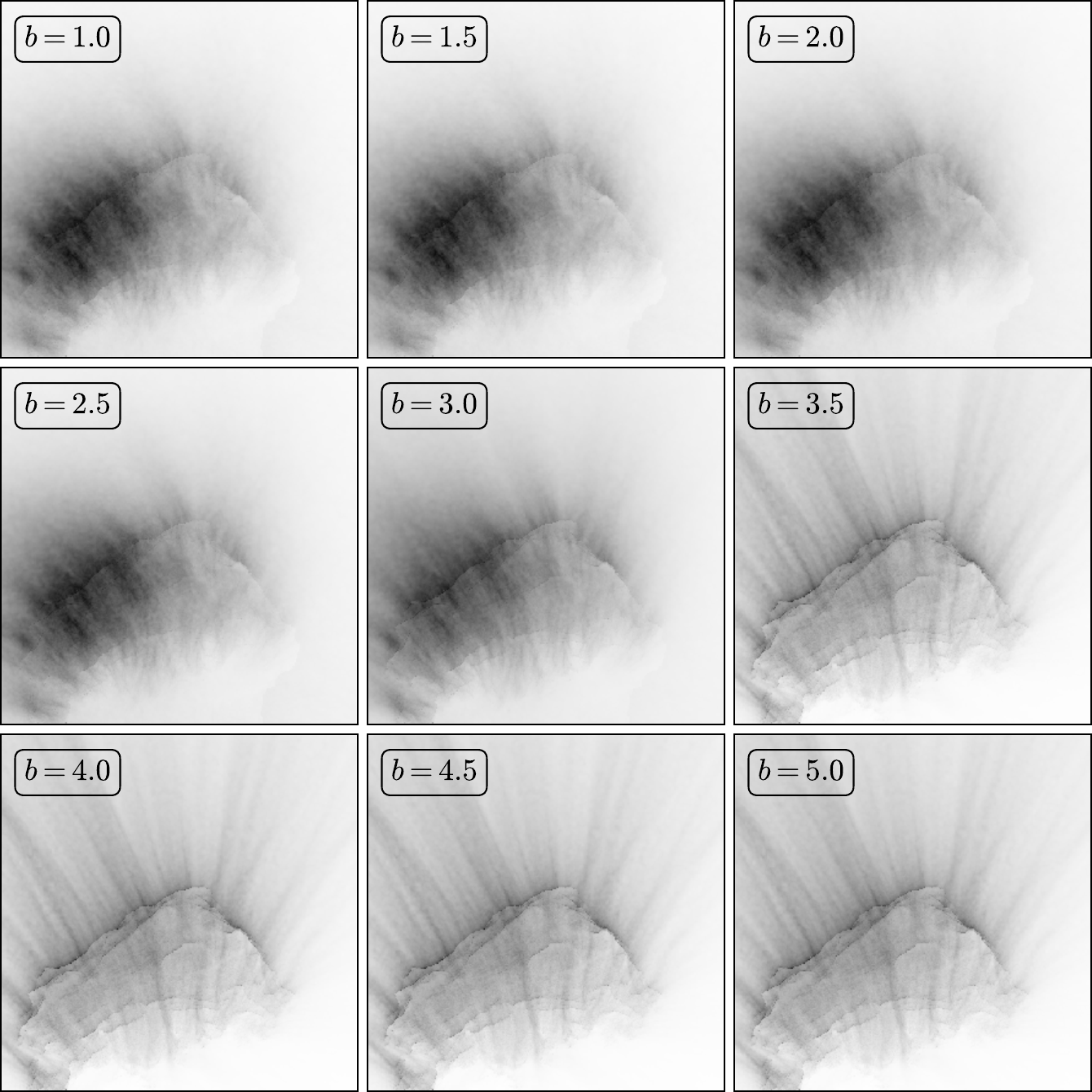}
                \caption{Global coma simulations for sequence STP087 as a function of the particle SFD power-law index, $b$, given $v_\text{init} = 0.5$\,m\,s$^{-1}$.}
                \label{fig:STP087_coma_simulations_global_vinit=0.50}
            \end{figure*}
            
            \begin{figure*}[!ht]
                \centering
                \includegraphics[width=0.59\linewidth]{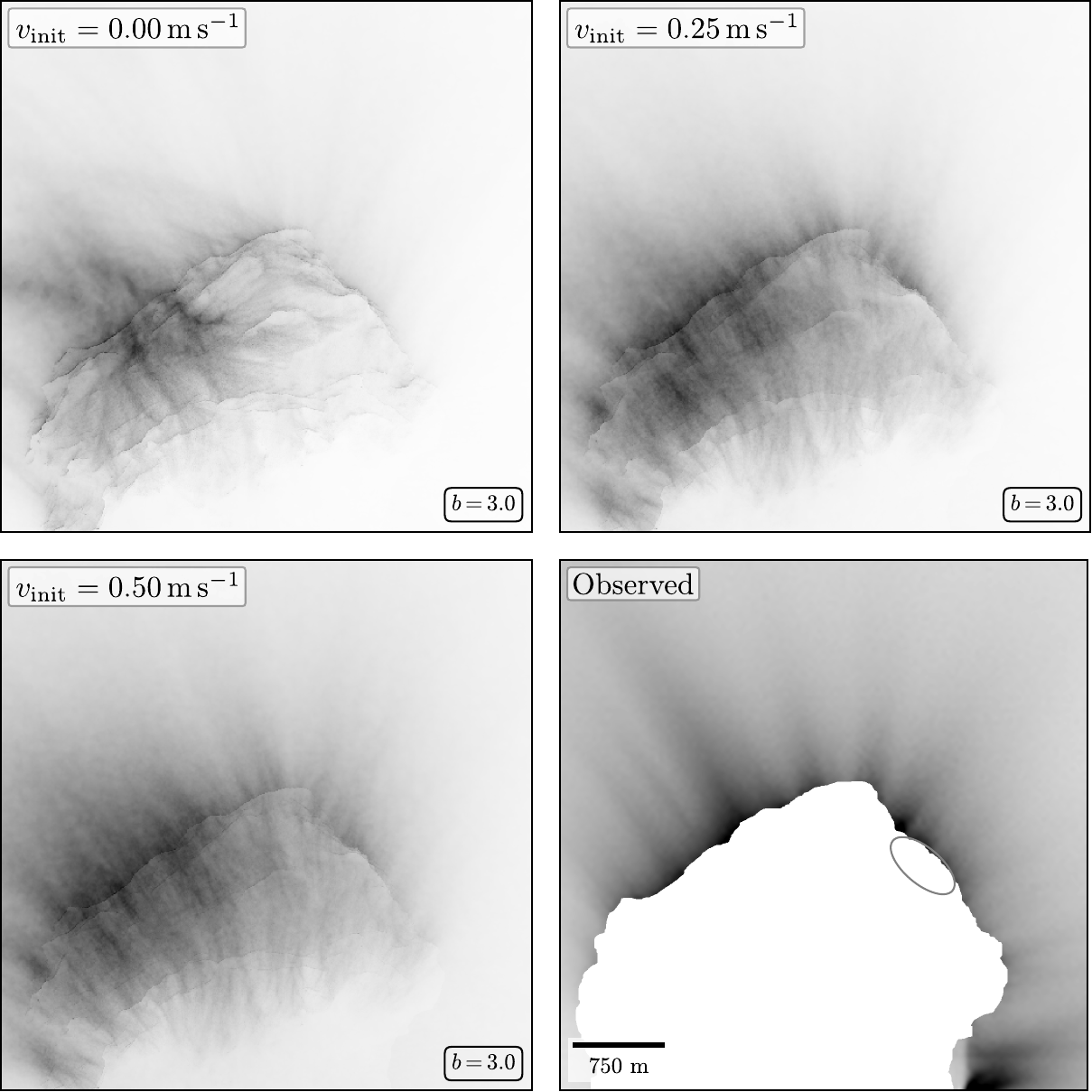}
                \caption{Comparison between global coma simulations given different initial velocities, $v_\text{init}$, and the observed coma in the first image of sequence STP087. The image of the observed coma is the background signal   subtracted during the preparation for the tracking procedure (see Sect.~\ref{sect:data_selection} and \citealp{pfeiferTrailCometTail2022}). The ellipse indicates the suspected source region. All images are brightness-inverted and had their contrasts improved individually for better readability. Because of this, the absolute intensity levels should not be compared across images, but only relative to other areas of the same image.}
                \label{fig:STP087_vinit_comparison}
            \end{figure*}

    \end{appendix}

\end{document}